\documentclass
[floatfix,superscriptaddress,secnumarabic,amssymb,amsmath,nobibnotes,aps,prd,showkeys,nofootinbib,onecolumn,12pt]{revtex4}%
\usepackage{setspace}
\usepackage{hyperref}
\usepackage{color}
\usepackage{amsmath}
\usepackage{amsfonts}
\usepackage{verbatim}
\usepackage{amssymb}
\usepackage{graphicx,bm}
\usepackage{graphicx}
\usepackage{amsmath}
\usepackage{amssymb}
\usepackage{amssymb}
\usepackage{graphicx,bm}
\usepackage{graphicx}
\usepackage[caption=false]{subfig}%
\providecommand{\U}[1]{\protect\rule{.1in}{.1in}}

\newcommand{\be}{\begin{equation}}
\newcommand{\ee}{\end{equation}}

\newcommand{\mincir}{\raise
-3.truept\hbox{\rlap{\hbox{$\sim$}}\raise4.truept\hbox{$<$}\ }}
\newcommand{\magcir}{\raise
-3.truept\hbox{\rlap{\hbox{$\sim$}}\raise4.truept\hbox{$>$}\ }}

\ifx\pdfoutput\relax\let\pdfoutput=\undefined\fi
\newcount\msipdfoutput
\ifx\pdfoutput\undefined\else
\ifcase\pdfoutput\else
\ifx\paperwidth\undefined\else
\ifdim\paperheight=0pt\relax\else\pdfpageheight\paperheight\fi
\ifdim\paperwidth=0pt\relax\else\pdfpagewidth\paperwidth\fi
\fi\fi\fi
\begin{document}
\title{Mapping Solutions in Nonmetricity Gravity: Investigating Cosmological Dynamics
in Conformal Equivalent Theories}

\author{Nikolaos Dimakis}
\email{nikolaos.dimakis@ufrontera.cl}
\affiliation{Departamento de Ciencias F\'{\i}sicas, Universidad de la Frontera, Casilla
54-D, 4811186 Temuco, Chile}

\author{Kevin Duffy}
\email{kevind@dut.ac.za}
\affiliation{Institute of Systems Science, Durban University of Technology, Durban 4000,
South Africa}

\author{Alex Giacomini}
\email{alexgiacomini@uach.cl }
\affiliation{Instituto de Ciencias F\'{\i}sicas y Matem\'{a}ticas, Universidad Austral de
Chile, Valdivia 5090000, Chile}

\author{Alexander Yu. Kamenshchik}
\email{Alexander.Kamenshchik@bo.infn.it}
\affiliation{Dipartimento di Fisica e Astronomia, Universit`a di Bologna, via Irnerio 46,
40126 Bologna, Italy}
\affiliation{I.N.F.N., Sezione di Bologna, I.S. FLAG, viale Berti Pichat 6/2, 40127
Bologna, Italy}

\author{Genly Leon}
\email{genly.leon@ucn.cl}
\affiliation{Departamento de Matem\'{a}ticas, Universidad Cat\'{o}lica del Norte, Avda.
Angamos 0610, Casilla 1280 Antofagasta, Chile}
\affiliation{Institute of Systems Science, Durban University of Technology, Durban 4000,
South Africa}

\author{Andronikos Paliathanasis}
\email{anpaliat@phys.uoa.gr}
\affiliation{Institute of Systems Science, Durban University of Technology, Durban 4000,
South Africa}
\affiliation{Departamento de Matem\'{a}ticas, Universidad Cat\'{o}lica del Norte, Avda.
Angamos 0610, Casilla 1280 Antofagasta, Chile}

\begin{abstract}
We investigate the impact of conformal transformations on the physical
properties of solution trajectories in nonmetricity gravity. Specifically, we
explore the phase space and reconstruct the cosmological history of a
spatially flat Friedmann--Lema\^{\i}tre--Robertson--Walker universe within
scalar-nonmetricity theory in both the Jordan and Einstein frames. A detailed
analysis is conducted for three connections defined in the
coincident and non-coincident gauges. Our findings reveal the existence of a
unique one-to-one correspondence for equilibrium points in the two frames.
Furthermore, we demonstrate that solutions describing accelerated universes
remain invariant under the transformation that relates these conformally
equivalent theories.

\end{abstract}
\keywords{Symmetric teleparallel; non-metricity gravity; scalar-tensor cosmology;
conformal transformation; phase-space analysis}\maketitle

\section{Introduction}

\label{sec1}

Conformal transformations are mappings that reshape geometric objects into
other forms, wherein the distances between points are not conserved. However,
these transformations maintain the angles at each point on the object. A
subset of conformal transformations is known as isometries, wherein
the distances between points remain unchanged. Isometries and conformal
transformations have various applications in different branches of physical
science, offering a systematic approach to analyze the physical world
\cite{conf}.

In Newtonian physics, isometries are instrumental in
understanding the conservation laws governing momentum and angular momentum
as they pertain to Euclidean geometry \cite{con1}. Similarly, in gravitational
physics, the conservation laws associated with time-like geodesics are related
to the presence of isometries for the background geometry. On the other
hand, conformal transformations are used to construct conservation laws for
the null-geodesics \cite{con2}.

Conformal transformations are crucial in 
gravitational physics and cosmology framework, particularly in scalar-tensor theories
\cite{ref1,ref2}. These transformations can transit between
the Jordan and Einstein frames and vice versa. In scalar-tensor
theories, a scalar field is nonminimally coupled to gravity, introducing a
coupling function in the gravitational Lagrangian to describe the interaction
with the scalar field \cite{ref3}. Through the application of a conformal
transformation, the coupling function can be eliminated from the gravitational
Lagrangian. This results in an equivalent theory where the gravitational
dynamics involve a minimally coupled scalar field defined in the Einstein
frame. It is important to note that introducing the coupling function
to the metric tensor through the conformal transformation leads to differences
in the physical quantities derived from the same solution trajectories
\cite{ref4}.

This mathematical approach enables the construction of new solutions for
conformally equivalent theories. Additionally, researchers have demonstrated
that applying conformal transformations can be instrumental in
avoiding cosmological singularities \cite{kam1}. Indeed, singular solutions in
the one frame can correspond to nonsingular solutions for the other frame and
vice versa \cite{ll1,ll2,ll3}. Thus, conformal transformations are a powerful
tool for understanding the dynamics of scalar-tensor theories in gravitational
physics and cosmology. Numerous studies have extensively investigated physical
quantities within the context of exact solutions in both the Einstein and
Jordan frames \cite{jd1,jd2,jd3,jd4,jd5,jd6}. Despite this wealth of research,
the question of which frame is preferred remains unanswered \cite{fr1}.

In this study, we deal with the effects of conformal transformations on the
physical properties of conformal equivalent theories in scalar-tensor theories
in the framework of symmetric teleparallel gravity. This theory, which from now
we will call it scalar-nonmetricity, which is an extension of Symmetric Teleparallel
General Relativity (STGR) \cite{Nester:1998mp} where a scalar field lies on
the physical space with a nonzero interaction function with the fundamental Lagrangian of the theory is the nonmetricity scalar $Q~$\cite{sc1}.
Scalar-nonmetricity theory is the analogue of the scalar-curvature \cite{sf1}
and scalar-torsion theories \cite{sf4} for the third invariant of the trinity
of gravity \cite{trinity}. STGR and its extensions
\cite{Koivisto2,Koivisto3,rev10} have been introduced as theoretical
frameworks aimed at addressing fundamental cosmological phenomena,
specifically the cosmic acceleration and the formation of the universe
\cite{ft1,ft2,ft3,ft5,ft6,fq2,fq3,fq4,fq6,fq8,fq9,fq10,fq11,fq12,fq14,fq15}.

In a recent work, \cite{bdpal}, a Brans-Dicke analogue was introduced within the
framework of symmetric teleparallel theory. This model, akin to the original
Brans-Dicke theory, introduced by \cite{Brans}, incorporates a free parameter
analogous to the Brans-Dicke parameter denoted as \cite{omegaBDGR}. Notably,
when this parameter is set to zero, the model describes the $f\left(
Q\right)  $-theory. This construction parallels the way in which the
Brans-Dicke theory allocates degrees of freedom for the $f\left(  R\right)
$-theory of gravity \cite{s01}. For the Brans-Dicke analogue in STGR, new
cosmological solutions are determined \cite{bdpal}, and the impact of the conformal
transformation of the physical variables are examined. It was found that the
generic properties of exact solutions remain invariant under the conformal
transformation \cite{bdpal}.

To conduct a detailed analysis of the impact of conformal transformations on
the physical properties of solution trajectories in scalar-nonmetricity
theory, we focus on investigating the phase space for these trajectories
within the context of a spatially flat
Friedmann-Lema\^{\i}tre-Robertson-Walker (FLRW) universe. Our specific goal is
to reconstruct the cosmological history as conformally equivalent theories describe, allowing us to compare the cosmic evolution
and relevant cosmological epochs. The structure of the paper is as follows.

In Section \ref{sec2}, we provide the fundamental properties and definitions
of scalar-nonmetricity gravity. This theory is a generalization of
scalar-curvature theories within the symmetric teleparallel formalism, where
the scalar field is nonminimally coupled to the nonmetricity scalar $Q$.
Additionally, we go deeper into the effects of conformal transformations and
determine the conformal equivalent theory. Our focus centers on the
Brans-Dicke analogue within nonmetricity gravity. In Section \ref{sec3}, we
specifically concentrate on this Brans-Dicke analogue. Here, we present the
field equations applicable to a spatially flat
Friedmann--Lema\^{\i}tre--Robertson--Walker (FLRW) geometry for three distinct
sets of the connection. Demonstrating that these field equations permit a
minisuperspace description, we proceed to formulate the corresponding
point-like Lagrangians for the nonmetricity Brans-Dicke model in both the
Jordan and \ (pseudo-) Einstein frames. As we shall see in the following, although
the conformal equivalent theory of the scalar field is not coupled to
the nonmetricity Lagrangian, there exists a nontrivial coupling function with
another geometric invariant related to the Lagrangian of the nonmetricity
theory. Hence, we shall call that the theory is defined in the a (pseudo-)
Einstein frame.

The phase-space analysis of the field equations and the reconstruction of the
cosmological history are presented in the respective Sections \ref{sec4},
\ref{sec5} and \ref{sec6} for the three different connections. Specifically,
we employ dimensionless variables within the $H$-normalization approach to
determine equilibrium points for the field equations. Our analysis extends to
investigating the physical properties of asymptotic solutions at these
equilibrium points and their stability properties. The
insights gained from this analysis are then utilized to define constraints for
the theory's viability. Furthermore, a similar
analysis is conducted for the conformal equivalent theory defined in the
Einstein frame to explore the impact of conformal transformations on the physical properties of solution trajectories. This comparative analysis reveals a one-to-one correspondence
between equilibrium points and their associated physical properties in both
theories. Finally, we present our conclusions in Section \ref{sec8}.

\section{Scalar-nonmetricity theory}

\label{sec2}

We consider the scalar-nonmetricity theory of gravity described by the Action
Integral \cite{sc1}
\begin{equation}
S_{ST\varphi}=\int d^{4}x\sqrt{-g}\left(  \frac{F\left(  \varphi\right)  }%
{2}Q+\frac{\omega\left(  \varphi\right)  }{2}g^{\mu\nu}\varphi_{,\mu}%
\varphi_{,\nu}+V\left(  \varphi\right)  \right), \label{ac.01}%
\end{equation}
where $\phi$ is a scalar field with potential function $V\left(  \phi\right)
$, $g_{\mu\nu}$ is the metric tensor of a four-dimensional manifold $M$ with
the symmetric connection $\Gamma_{\mu\nu}^{\lambda}$ which inherits the
symmetries of the metric tensor $g_{\mu\nu}$ and defined the covariant
derivative operator $\nabla_{\lambda}$.

The gravitational scalar $Q,$ which is is the nonmetricity scalar, is defined
as \cite{rev10}
\begin{equation}
Q=Q_{\lambda\mu\nu}P^{\lambda\mu\nu}%
\end{equation}
where $Q_{\lambda\mu\nu}=\nabla_{\lambda}g_{\mu\nu}$ is the nonmetricity
tensor, that is,%
\begin{equation}
Q_{\lambda\mu\nu}=\frac{\partial g_{\mu\nu}}{\partial x^{\lambda}}%
-\Gamma_{\;\lambda\mu}^{\sigma}g_{\sigma\nu}-\Gamma_{\;\lambda\nu}^{\sigma
}g_{\mu\sigma}.
\end{equation}

The geometric object $P^{\lambda\mu\nu}$ is the nonmetricity conjugate tensor
\cite{Hohmann}
\begin{equation}
P_{~\mu\nu}^{\lambda}=\frac{1}{4}\left(  -2L_{~~\mu\nu}^{\lambda}+Q^{\lambda
}g_{\mu\nu}-Q^{\prime\lambda}g_{\mu\nu}-\delta_{(\mu}^{\lambda}Q_{\nu
)}\right), 
\end{equation}
where%
\begin{equation}
L_{~\mu\nu}^{\lambda}=\frac{1}{2}g^{\lambda\sigma}\left(  Q_{\mu\nu\sigma
}+Q_{\nu\mu\sigma}-Q_{\sigma\mu\nu}\right)
\end{equation}
and%
\[
Q_{\lambda}=Q_{\lambda~~~\mu}^{~~~\mu},Q_{\lambda}^{\prime}=Q_{~~\lambda\mu
}^{\mu}~.
\]

Furthermore, function $F\left(  \phi\right)  $ in (\ref{ac.01}) is the
coupling function between the scalar field and the nonmetricity scalar;
similarly to the coupling function of the scalar field with the Ricci scalar
in the scalar-curvature theory. On the other hand, function $\omega\left(
\phi\right)  $ can be eliminated with the introduction of the new scalar field
$d\Phi=\sqrt{\omega\left(  \varphi\right)  }d\varphi$; where Action
(\ref{ac.01}) reads \cite{sc1}%
\begin{equation}
S_{ST\Phi}=\int d^{4}x\sqrt{-g}\left(  \frac{F\left(  \Phi\right)  }{2}%
Q+\frac{1}{2}g^{\mu\nu}\Phi_{,\mu}\Phi_{,\nu}+V\left(  \Phi\right)  \right). 
\end{equation}

Variation with respect to the metric tensor in (\ref{ac.01}) leads to the
field equations \cite{sc1,sc2}%
\begin{equation}
F\left(  \varphi\right)  G_{\mu\nu}+2F_{,\phi}\varphi_{,\lambda}P_{~~\mu\nu
}^{\lambda}-g_{\mu\nu}V\left(  \varphi\right)  -\frac{\omega\left(
\varphi\right)  }{2}\left(  g_{\mu\nu}g^{\lambda\kappa}\varphi_{,\lambda
}\varphi_{,\kappa}-\varphi_{,\mu}\varphi_{,\nu}\right)  =0,
\end{equation}
while variation with respect to the connection $\Gamma_{\mu\nu}^{\lambda}$
leads to the equations%
\begin{equation}
\nabla_{\mu}\nabla_{\nu}\left(  \sqrt{-g}F\left(  \varphi\right)
P_{\phantom{\mu\nu}\sigma}^{\mu\nu}\right)  =0. \label{cc1}%
\end{equation}
Finally, variation with respect to the scalar field in (\ref{ac.01}) provides
the modified Klein-Gordon equation%
\begin{equation}
\frac{\omega\left(  \varphi\right)  }{\sqrt{-g}}g^{\mu\nu}\partial_{\mu
}\left(  \sqrt{-g}\partial_{\nu}\varphi\right)  +\frac{\omega_{,\varphi}}%
{2}g^{\lambda\kappa}\varphi_{,\lambda}\varphi_{,\kappa}-\frac{1}{2}%
F_{,\varphi}Q-V_{,\varphi}=0.
\end{equation}

It is important to observe that for $\omega\left(  \varphi\right)  =0$,
$F\left(  \varphi\right)  =\varphi$, the latter field equations take the
functional form of $f\left(  Q\right)  $-theory \cite{sc1}, where now
$\varphi=f^{\prime}\left(  Q\right)  $ and $V\left(  \varphi\right)  =\left(
f^{\prime}(Q)Q-f(Q)\right)  $, which means that the Action (\ref{ac.01}) is
equivalent to that of $f\left(  Q\right)  $-theory \cite{Koivisto2,Koivisto3}.

\subsection{Conformal equivalent theory}

\label{sec2a}

Let $\bar{g}_{\mu\nu},~g_{\mu\nu}$ be two metric tensors that share the same
conformal algebra, meaning that the metrics are conformally related in such a
way that
\[
\bar{g}_{\mu\nu}=e^{2\Omega\left(  x^{\kappa}\right)  }g_{\mu\nu}~,~~\bar
{g}^{\mu\nu}=e^{-2\Omega\left(  x^{\kappa}\right)  }g^{\mu\nu}.
\]
where $\Omega\left(  x^{\kappa}\right)  $ is the so-called conformal function.

The nonmetricity tensors $\bar{Q}_{\lambda\mu\nu}$, $Q_{\lambda\mu\nu},$ for
the two conformal related metrics are related as \cite{gg1}%
\begin{equation}
\bar{Q}_{\lambda\mu\nu}=e^{2\Omega}Q_{\lambda\mu\nu}+2\Omega_{,\lambda}\bar
{g}_{\mu\nu}.
\end{equation}
and the corresponding nonmetricity scalars $\bar{Q}$ and $Q$ are related%
\begin{equation}
\bar{Q}=e^{-2\Omega}Q+\left(  2\Omega_{,\lambda}P^{\lambda}+6\Omega_{\lambda
}\Omega^{,\lambda}\right). 
\end{equation}

Assume the Action Integral (\ref{ac.01}) for the metric tensor $\bar{g}%
_{\mu\nu}$, that is,%
\begin{equation}
\bar{S}_{ST\varphi}=\int d^{4}x\sqrt{-\bar{g}}\left(  \frac{F\left(
\varphi\right)  }{2}\bar{Q}+\frac{\omega\left(  \varphi\right)  }{2}\bar
{g}^{\mu\nu}\varphi_{,\mu}\varphi_{,\nu}+V\left(  \varphi\right)  \right). 
\end{equation}

Then, the conformal equivalent theory is
\begin{equation}
\bar{S}_{ST\varphi}=\int d^{4}x\sqrt{-g}\left(  \frac{Q}{2}-\ln F\left(
\varphi\right)  \frac{B}{4}+\left(  \frac{3\left(  F_{,\varphi}\right)
^{2}+2\omega\left(  \varphi\right)  }{2F\left(  \varphi\right)  }\right)
g^{\mu\nu}\varphi_{,\mu}\varphi_{,\nu}+\frac{V\left(  \varphi\right)
}{\left(  F\left(  \phi\right)  \right)  ^{2}}\right), 
\end{equation}
where $B=\mathring{R}-Q,$ is the boundary term relates the nonmetricity scalar
$Q$ and the Ricci scalar $\mathring{R}$ for the Levi-Civita connection
$\mathring{\Gamma}_{~\mu\nu}^{\lambda}$of the metric tensor $g_{\mu\nu}$
\cite{sc1}.

We introduce the new scalar field
\begin{equation}
d\Psi=\sqrt{\frac{3\left(  F_{,\varphi}\right)  ^{2}-2\omega\left(
\varphi\right)  }{2F\left(  \varphi\right)  }}d\varphi,
\end{equation}
and the latter Action Integral becomes%
\begin{equation}
\bar{S}_{ST\Psi}=\int d^{4}x\sqrt{-g}\left(  \frac{Q}{2}-\ln F\left(
\Psi\right)  \frac{B}{4}+g^{\mu\nu}\Psi_{,\mu}\Psi_{,\nu}+\tilde{V}\left(
\Psi\right)  \right), ~~\tilde{V}\left(  \Psi\right)  =\frac{V\left(
\Psi\right)  }{\left(  F\left(  \Psi\right)  \right)  ^{2}}.
\end{equation}

\subsection{Nonmetricity Brans-Dicke theory}

We consider the scalar-nonmetricity theory with $F\left(  \varphi\right)
=\varphi$ and $\omega\left(  \varphi\right)  \rightarrow-\frac{\omega}%
{\varphi}$ in which$~\omega=const.$ This theory can be seen as the extension
of the Brans-Dicke theory in nonmetricity scalar, where $\omega$ plays the role of
the Brans-Dicke parameter. Indeed, the Action Integral (\ref{ac.01}) reads
\cite{gg1,bdpal}%

\begin{equation}
S_{BD\varphi}=\int d^{4}x\sqrt{-g}\left(  \frac{\varphi}{2}Q+\frac{\omega
}{2\varphi}g^{\mu\nu}\varphi_{,\mu}\varphi_{,\nu}+V\left(  \varphi\right)
\right). 
\end{equation}

An equivalent way to write the latter theory is by introducing the dilaton
field $\varphi=e^{\phi}$, such that the latter Action Integral is
\begin{equation}
S_{D}=\int d^{4}x\sqrt{-g}e^{\phi}\left(  \frac{Q}{2}+\frac{\omega}{2}%
g^{\mu\nu}\phi_{,\mu}\phi_{,\nu}+\hat{V}\left(  \phi\right)  \right)
,~~\hat{V}\left(  \phi\right)  =V\left(  \phi\right)  e^{-\phi}.\label{sd.01}%
\end{equation}

Moreover, the action integral for the conformally equivalent theory is given
by
\begin{equation}
\bar{S}_{D}=\int d^{4}x\sqrt{-g}\left(  \frac{Q}{2}-\phi\frac{B}{4}+\frac
{\bar{\omega}}{2}g^{\mu\nu}\phi_{,\mu}\phi_{,\nu}+V\left(  \phi\right)
e^{-2\phi}\right), ~\bar{\omega}=\frac{3}{2}+\omega,~~\bar{V}\left(
\phi\right)  =V\left(  \phi\right)  e^{-2\phi}.\label{sd.02}%
\end{equation}

\section{FLRW Cosmology}

\label{sec3}

In this study, we investigate the effects of conformal transformation in
the cosmological evolution and cosmological history. Specifically, we consider
a universe described by the isotropic and homogeneous FLRW geometry, with
element
\begin{equation}
ds^{2}=-N(t)^{2}dt^{2}+a(t)^{2}\left[  \frac{dr^{2}}{1-kr^{2}}+r^{2}\left(
d\theta^{2}+\sin^{2}\theta d\varphi^{2}\right)  \right], \label{genlineel}%
\end{equation}
in which $N\left(  t\right)  $ is the lapse function, $a\left(  t\right)  $ is
the scale factor denotes the radius of the universe. The Hubble function is
defined as, $H=\frac{1}{N}\frac{\dot{a}}{a}$, where $\dot{a}=\frac{da}{dt}$ .
$k$ denotes the spatial curvature, for $k=0$, the universe is spatially flat,
$k=+1$ corresponds to a closed FLRW\ geometry and $k=-1$ describes an open universe.

For this cosmological model, we study the dynamics of the field equations in
scalar-nonmetricity theory for the dilaton field (\ref{sd.01}) and will
reconstruct the cosmological history. Furthermore, we will perform the same
analysis for the conformal equivalent theory (\ref{sd.02}). We shall compare
the two cosmological histories and the provided cosmological eras by the two
different cosmological models. From this analysis, we can infer
the effects of the conformal transformation on the physical solutions in
nonmetricity theories.

In General Relativity, the definition of the connection is unique; it is the
Levi-Civita, in nonmetricity theory, the connection is not necessarily
unambiguously defined. For the FLRW geometry, four different
families of connections are used to describe diagonal field equations
\cite{Hohmann}. For the spatially flat universe, there are three different
families of connections; on the other hand, for $k\neq0$, the connection is
uniquely defined. In the following, we consider that the spatial curvature is zero.

For $k=0$, the common nonzero components of the of the three different
connections $\Gamma_{1}$,~$\Gamma_{2}$ and $\Gamma_{3}$ are \cite{Hohmann}%
\[
\Gamma_{\theta\theta}^{r}=-r,~~\Gamma_{\varphi\varphi}^{r}=-r\sin^{2}\theta
\]%
\[
\Gamma_{\varphi\varphi}^{\theta}=-\sin\theta\cos\theta,~~\Gamma_{\theta
\varphi}^{\varphi}=\Gamma_{\varphi\theta}^{\varphi}=\cot\theta
\]%
\[
\Gamma_{\;r\theta}^{\theta}=\Gamma_{\;\theta r}^{\theta}=\Gamma_{\;r\varphi
}^{\varphi}=\Gamma_{\;\varphi r}^{\varphi}=\frac{1}{r}%
\]
while the additional nonzero components for each connection $\Gamma_{1}%
$,~$\Gamma_{2}$ and $\Gamma_{3}$ are \cite{Hohmann}%
\[
\Gamma_{1}:\Gamma_{\;tt}^{t}=\gamma(t),
\]%
\[
\Gamma_{2}:\Gamma_{\;tt}^{t}=\frac{\dot{\gamma}(t)}{\gamma(t)}+\gamma
(t),\quad\Gamma_{\;tr}^{r}=\Gamma_{\;rt}^{r}=\Gamma_{\;t\theta}^{\theta
}=\Gamma_{\;\theta t}^{\theta}=\Gamma_{\;t\varphi}^{\varphi}=\Gamma_{\;\varphi
t}^{\varphi}=\gamma(t),
\]
and%
\[
\Gamma_{3}:\Gamma_{\;tt}^{t}=-\frac{\dot{\gamma}(t)}{\gamma(t)},\quad
\Gamma_{\;rr}^{t}=\gamma(t),\quad\Gamma_{\;\theta\theta}^{t}=\gamma
(t)r^{2},\quad\Gamma_{\;\varphi\varphi}^{t}=\gamma(t)r^{2}\sin^{2}\theta,
\]
where a dot means derivative with respect to the time parameter $t$, i.e.
$\dot{\gamma}=\frac{d\gamma}{dt}$.

Consequently, the nonmetricity scalars $Q$ and the corresponding boundary
terms $B$ for the first connection $\Gamma_{1}$ read \cite{anbb}%
\begin{equation}
Q_{1}\left(  \Gamma_{1}\right)  =-6H^{2},~
\end{equation}%
\begin{equation}
B_{1}\left(  \Gamma_{1}\right)  =3\left(  6H^{2}+\frac{2}{N}\dot{H}\right), ~
\end{equation}
for the second connection $\Gamma_{2}$ are calculated \cite{anbb}%
\begin{equation}
Q_{2}\left(  \Gamma_{2}\right)  =-6H^{2}+\frac{3}{a^{3}N}\left(  \frac
{a^{3}\gamma}{N}\right)  ^{\cdot},
\end{equation}%
\begin{equation}
B_{2}\left(  \Gamma_{2}\right)  =3\left(  6H^{2}+\frac{2}{N}\dot{H}-\frac
{3}{a^{3}N}\left(  \frac{a^{3}\gamma}{N}\right)  ^{\cdot}\right). 
\end{equation}
while for the third connection $\Gamma_{3}$ we calculate the scalars
\cite{anbb}%
\begin{equation}
Q_{3}\left(  \Gamma_{3}\right)  =-6H^{2}+\frac{3}{a^{3}N}\left(
aN\gamma\right)  ^{\cdot},
\end{equation}%
\begin{equation}
B_{3}\left(  \Gamma_{3}\right)  =3\left(  6H^{2}+\frac{2}{N}\dot{H}-\frac
{1}{a^{3}N}\left(  aN\gamma\right)  ^{\cdot}\right). 
\end{equation}

\subsection{Minisuperspace description for the dilaton field}

For each connection the resulting field equations are different. That is,
because the coupling between the scalar field $\phi$ with the nonmetricity
scalar $Q$ leads to the introduction of dynamical degrees of freedom related
to the function $\gamma\left(  t\right)  $ which defines the connection.
Connection $\Gamma_{1}$ is defined in the so-called coincident gauge where the
equation of motion (\ref{cc1}) is trivially satisfied. However, that is not
true for the other three families of connections that are defined in the
noncoincident gauge.

To understand the effects of the connection in the field equations, we
follow the procedure described in \cite{mini} and we write the\ corresponding
point-like Lagrangian for the field equations for each connection.

For the first connection, namely $\Gamma_{1}$, the corresponding point-like
Lagrangian is
\begin{equation}
L\left(  \Gamma_{1}\right)  =\frac{e^{\phi}}{N}\left(  3a\dot{a}^{2}%
+\frac{\omega}{2}a^{3}\dot{\phi}^{2}\right)  -Na^{3}V\left(  \phi\right). 
\label{rs.01}%
\end{equation}
Similarly, for the second connection, $\Gamma_{2}$, the field equations follow
from the variation of the point-like Lagrangian function \cite{anbb}%
\begin{equation}
L\left(  \Gamma_{2}\right)  =\frac{e^{\phi}}{N}\left(  3a\dot{a}^{2}%
+\frac{\omega}{2}a^{3}\dot{\phi}^{2}+\frac{3}{2}a^{3}\dot{\phi}\dot{\psi
}\right)  -Na^{3}V\left(  \phi\right),  \label{rs.02}%
\end{equation}
in which $\gamma\left(  t\right)  =\dot{\psi}\left(  t\right)  $.

Finally, for connection $\Gamma_{3}$ and $\Gamma_{4}$ the Lagrangian function
is
\begin{equation}
L_{3}\left(  \Gamma_{3}\right)  =\frac{e^{\phi}}{N}\left(  3a\dot{a}^{2}%
+\frac{\omega}{2}a^{3}\dot{\phi}^{2}+\frac{3}{2}aN^{2}\frac{\dot{\phi}}%
{\dot{\Psi}}\right)  -Na^{3}V\left(  \phi\right).  \label{rs.03}%
\end{equation}

\subsubsection{Conformal transformation}

We can also write the minisuperspace Lagrangians and the conformal
equivalent theories. Indeed, the FLRW\ line element%

\begin{equation}
d\bar{s}^{2}=-\bar{N}^{2}\left(  t\right)  dt^{2}+\alpha^{2}\left(  t\right)
\left(  dr^{2}+r^{2}\left(  d\theta^{2}+\sin^{2}\theta d\phi^{2}\right)
\right), \label{rs.04}%
\end{equation}
with $a\left(  t\right)  =\alpha\left(  t\right)  e^{-\frac{\phi\left(
t\right)  }{2}}$, $N\left(  t\right)  =\bar{N}\left(  t\right)  e^{-\frac
{\phi\left(  t\right)  }{2}}$ is conformally related to the line element
(\ref{genlineel}) with conformal factor the coupling function $e^{-\phi\left(
t\right)  }$.

By applying the latter transformation in the action
\begin{equation}
S=\int L\left(  \mathbf{\Gamma}\right)  dt\text{, }%
\end{equation}
for each of the Lagrangian functions (\ref{rs.01}), (\ref{rs.02}) and
(\ref{rs.03}); we end with the following conformal equivalent point-like Lagrangians%

\begin{equation}
\bar{L}_{1}\left(  \Gamma_{1}\right)  =\frac{1}{\bar{N}}\left(  3\alpha
\dot{\alpha}^{2}-3\alpha^{2}\dot{\alpha}\dot{\phi}+\frac{\bar{\omega}}%
{2}\alpha^{3}\dot{\phi}^{2}\right)  -\bar{N}\alpha^{3}\bar{V}\left(
\phi\right), ~~ \label{rs.05}%
\end{equation}%
\begin{equation}
\bar{L}_{2}\left(  \Gamma_{2}\right)  =\frac{1}{\bar{N}}\left(  3\alpha
\dot{\alpha}^{2}-3\alpha^{2}\dot{\alpha}\dot{\phi}+\frac{\bar{\omega}}%
{2}\alpha^{3}\dot{\phi}^{2}+\frac{3}{2}\alpha^{3}\dot{\phi}\dot{\psi}\right)
-\bar{N}\alpha^{3}\bar{V}\left(  \phi\right),  \label{rs.06}%
\end{equation}
and%
\begin{equation}
\bar{L}_{3}\left(  \Gamma_{3}\right)  =\frac{1}{\bar{N}}\left(  3\alpha
\dot{\alpha}^{2}-3\alpha^{2}\dot{\alpha}\dot{\phi}+\frac{\bar{\omega}}%
{2}\alpha^{3}\dot{\phi}^{2}+\frac{3}{2}\alpha N^{2}\frac{\dot{\phi}}{\dot
{\Psi}}\right)  -\bar{N}\alpha^{3}\bar{V}\left(  \phi\right).  \label{rs.07}%
\end{equation}
where now $\bar{V}\left(  \phi\right)  =e^{-2\phi}V\left(  \phi\right)
,~~\bar{\omega}=\omega+\frac{3}{2}.$

We remark that the conformal transformation eliminates the coupling function
$e^{\phi}$ in the Lagrangian, however, introduces the dynamical components
$-3\alpha^{2}\dot{\alpha}\dot{\phi}+\frac{3}{2}\alpha^{3}\dot{\phi}^{2}$ in
all the set of field equations. Moreover, for the scalar field potential
$V\left(  \phi\right)  $ we consider the exponential function, that is,
$V\left(  \phi\right)  =V_{0}e^{\lambda\phi}$.

\section{Phase-space analysis for connection $\Gamma_{1}$}

\label{sec4}

For the connection $\Gamma_{1}$ defined in the coincident gauge and for
$N\left(  t\right)  =1$, and from the point-like Lagrangian (\ref{rs.01}) we
determine the cosmological field equations, $\omega=cons\not t,$
\begin{align}
3H^{2}+\frac{\omega}{2}\dot{\phi}^{2}+e^{-\phi}V\left(  \phi\right)   &  =0,\\
2\dot{H}+3H^{2}+2H\dot{\phi}-\frac{\omega}{2}\dot{\phi}^{2}+e^{-\phi}V\left(
\phi\right)   &  =0,\\
3H^{2}-3\omega H\dot{\phi}-\frac{\omega}{2}\left(  \dot{\phi}^{2}+2\ddot{\phi
}\right)  -e^{-\phi}V_{,\phi}\left(  \phi\right)   &  =0.
\end{align}
where $H=\frac{\dot{a}}{a}$ is the Hubble function.

The latter equations can be written in the equivalent form
\begin{equation}
3H^{2}=\rho_{eff}\left(  \Gamma_{1}\right), ~~-2\dot{H}-3H^{2}=p_{eff}\left(
\Gamma_{1}\right)
\end{equation}
in which $\rho_{eff}$, $p_{eff}$ are the effective fluid energy density and
pressure component for the geometric fluid, defined
\begin{align}
\rho_{eff}\left(  \Gamma_{1}\right)   &  =-\left(  \frac{\omega}{2}\dot{\phi
}^{2}+e^{-\phi}V\left(  \phi\right)  \right), \\
p_{eff}\left(  \Gamma_{1}\right)   &  =2H\dot{\phi}-\frac{\omega}{2}\dot{\phi
}^{2}+e^{-\phi}V\left(  \phi\right), 
\end{align}

To examine the cosmological dynamics and reconstruct the cosmological
history for this gravitational model we introduce dimensionless variables in
the $H$-normalization consideration.

We define the new dependent variables
\begin{equation}
x=\frac{\dot{\phi}}{\sqrt{6}H},~~y=\frac{e^{-\phi}V}{3H^{2}},~
\end{equation}
and the independent variable $\tau=\ln a.$

Hence, the field equations with the use of the dimensionless variables are
expressed as follow%
\begin{align}
\frac{dx}{d\tau}  &  =\frac{x}{2}\left(  3\left(  y-\omega x^{2}-1\right)
+\sqrt{6}x\right)  +\frac{\sqrt{6}}{2\omega}\left(  1-\lambda y\right)
,\label{ds.01}\\
\frac{dy}{d\tau}  &  =y\left(  \sqrt{6}\left(  1+\lambda\right)  x-3\left(
\omega x^{2}-1-y\right)  \right),  \label{ds.02}%
\end{align}
with constraint equation%
\begin{equation}
1+\omega x^{2}+y=0. \label{ds.03}%
\end{equation}

Furthermore, the equation of state parameter is expressed as
\begin{equation}
w^{\Gamma_{1}}\left(  x,y\right)  =x\left(  \sqrt{\frac{8}{3}}-\omega
x\right)  +y. \label{ds.04}%
\end{equation}

With the application of the constraint equation (\ref{ds.03}) we can reduce
the dimension of the dynamical system (\ref{ds.01}), (\ref{ds.02}) by one.
Thus we end with the equation%
\begin{equation}
\frac{dx}{d\tau}=\frac{1}{2\omega}\left(  1+\omega x^{2}\right)  \left(
\sqrt{6}\left(  1+\lambda\right)  -6\omega x\right).  \label{ds.05}%
\end{equation}

The equilibrium points of the latter equation are
\[
A_{1}=\frac{\sqrt{6}\left(  1+\lambda\right)  }{6\omega},~~A_{2}^{\pm}%
=\frac{1}{\sqrt{-\omega}}.
\]

Point $A_{1}$ exist for $\omega\neq0$, and describe a universe dominated by a
fluid source with the equation of state parameter $w^{\Gamma_{1}}\left(
A_{1}\right)  =-1+\frac{1-\lambda^{2}}{3\omega}$. The latter asymptotic
solution describes a de Sitter universe for $\lambda^{2}=1$. On the other
hand, points $A_{2}^{\pm}$ are real for $\omega<0$. The equation of state
parameters for the asymptotic solutions at these two points are $w^{\Gamma
_{1}}\left(  A_{2}^{\pm}\right)  =1\pm\sqrt{\frac{8}{3\left\vert
\omega\right\vert }}$. Thus $w^{\Gamma_{1}}\left(  A_{2}^{+}\right)  >1$ and
$w^{\Gamma_{1}}\left(  A_{2}^{-}\right)  <1$. Hence, point $A_{2}^{-}$
describes an accelerated universe for $\left\vert \omega\right\vert <\frac
{3}{2}$. \

In order to investigate the stability properties of the linearized system we
calculate the eigenvalues of the linearized equation (\ref{ds.05}). They are
$e\left(  A_{1}\right)  =-3-\frac{\left(  1+\lambda\right)  ^{2}}{2\omega}$,
$e\left(  A_{2}^{\pm}\right)  =\sqrt{6}\left(  \sqrt{6}\pm\frac{\left(
1+\lambda\right)  }{\sqrt{\left\vert \omega\right\vert }}\right)  $. Thus,
point $A_{1}$ is an attractor for $\omega>0$ or $\omega<-\frac{\left(
1+\lambda\right)  ^{2}}{6}$. Furthermore, point $A_{2}^{+}$ is attractor for
$\lambda<-1$ and $\left\vert \omega\right\vert <\frac{\left(  1+\lambda
\right)  ^{2}}{6}$ and $A_{2}^{-}$ is attractor for $\lambda>-1$ and
$\left\vert \omega\right\vert <\frac{\left(  1+\lambda\right)  ^{2}}{6}$.

\subsection{Conformal equivalent theory}

Consider now the field equations for the conformal equivalent theory described
by the Lagrangian function (\ref{rs.05}), the equations are%
\begin{align}
3\bar{H}^{2}-3\bar{H}\dot{\phi}+\frac{\bar{\omega}}{2}\dot{\phi}^{2}+\bar
{V}\left(  \phi\right)   &  =0,\\
2\left(  \bar{H}\right)  ^{\cdot}+3\bar{H}^{2}-\frac{\bar{\omega}}{2}\dot
{\phi}^{2}-\ddot{\phi}+\bar{V}\left(  \phi\right)   &  =0,\\
\bar{\omega}\left(  \ddot{\phi}+3\bar{H}\dot{\phi}\right)  -3\left(  \left(
\bar{H}\right)  ^{\cdot}+3H^{2}\right)  +\bar{V}_{,\phi}\left(  \phi\right)
&  =0,
\end{align}
where we have assumed $\bar{N}\left(  t\right)  =1$.

Equivalently they can be expressed
\begin{equation}
3\bar{H}^{2}=\bar{\rho}_{eff}\left(  \Gamma_{1}\right), ~~-2\left(  \bar
{H}\right)  ^{\cdot}-3\bar{H}^{2}=\bar{p}_{eff}\left(  \Gamma_{1}\right), 
\end{equation}
with effective fluid components%
\begin{align}
\bar{\rho}_{eff}\left(  \Gamma_{1}\right)   &  =3\bar{H}\dot{\phi}-\left(
\frac{\bar{\omega}}{2}\dot{\phi}^{2}+\bar{V}\left(  \phi\right)  \right), \\
\bar{p}_{eff}\left(  \Gamma_{1}\right)   &  =-\left(  \frac{\bar{\omega}}%
{2}\dot{\phi}^{2}-\bar{V}\left(  \phi\right)  \right)  -\ddot{\phi},
\end{align}
in which $\bar{H}=\frac{\dot{\alpha}}{\alpha}$ is the Hubble function for the
conformal equivalent theory.

We follow the same procedure as before and we introduce the dimensionless
variables%
\[
\bar{x}=\frac{\dot{\phi}}{\sqrt{6}\bar{H}},~~\bar{y}=\frac{\bar{V}\left(
\phi\right)  }{3\bar{H}^{2}},~~\bar{\tau}=\ln\alpha\text{.}%
\]

Hence, the field equations in the set of variables ~$\left(  \bar{x}\left(
\bar{\tau}\right), \bar{y}\left(  \bar{\tau}\right)  \right)  $ read%
\begin{align}
\frac{d\bar{x}}{d\bar{\tau}}  &  =\frac{\sqrt{6}\left(  3+\bar{y}\left(
1-2\lambda\right)  \right)  -3x\left(  2\left(  \bar{\omega}+3\right)
-3\sqrt{6}\bar{\omega}\bar{x}+2\bar{\omega}^{2}\bar{x}^{2}-2\left(
\bar{\omega}+\lambda-2\right)  \bar{y}\right)  }{2\left(  2\bar{\omega
}-3\right)  },\label{ds.06}\\
\frac{d\bar{y}}{d\bar{\tau}}  &  =\frac{\bar{y}}{2\bar{\omega}-3}\left(
\sqrt{6}\bar{x}\left(  3\left(  2-\lambda\right)  +2\bar{\omega}\left(
1+\lambda\right)  \right)  -6\bar{\omega}^{2}\bar{x}^{2}+6\left(  \bar{\omega
}-3+\left(  \bar{\omega}+\lambda-2\right)  y\right)  \right),  \label{ds.07}%
\end{align}
and algebraic constraint%
\begin{equation}
1-\sqrt{6}\bar{x}+\bar{\omega}\bar{x}^{2}+\bar{y}=0. \label{ds.08}%
\end{equation}

Furthermore, the equation of state parameter is defined as
\begin{equation}
\bar{w}^{\Gamma_{1}}\left(  \bar{x},\bar{y}\right)  =\frac{2\sqrt{6}%
\bar{\omega}\bar{x}-3-2\bar{\omega}^{2}\bar{x}^{2}+2\left(  \bar{\omega
}+\lambda-2\right)  \bar{y}}{2\bar{\omega}-3}. \label{ds.09}%
\end{equation}

With the application of the constraint (\ref{ds.08}) we end with the single
first-order ordinary differential equation%
\begin{equation}
\frac{d\bar{x}}{d\bar{\tau}}=\frac{\sqrt{6}\left(  1+\lambda\right)  +\bar
{x}\left(  \sqrt{6}\left(  7\bar{\omega}-6+\lambda\left(  \bar{\omega
}+3\right)  \right)  \bar{x}-3\bar{\omega}\left(  \lambda-2+2\bar{\omega
}\right)  \bar{x}^{2}-3\left(  2\bar{\omega}+3\lambda\right)  \right)  }%
{2\bar{\omega}-3}. \label{ds.10}%
\end{equation}

The stationary points of the latter equation are
\[
\bar{A}_{1}=\frac{\sqrt{6}\left(  1+\lambda\right)  }{3\left(  2\left(
\bar{\omega}-1\right)  +\lambda\right)  },~~\bar{A}_{2}^{\pm}=\frac{\sqrt
{6}\pm\sqrt{2\left(  3-2\bar{\omega}\right)  }}{2\bar{\omega}}.
\]
or%
\[
\bar{A}_{1}=\frac{\sqrt{6}\left(  1+\lambda\right)  }{3\left(  2\omega
+\lambda+1\right)  },~~\bar{A}_{2}^{\pm}=\frac{\sqrt{6}\pm2\sqrt{-\omega}%
}{3+2\omega}.
\]

The equilibrium point $\bar{A}_{1}$ exist always and describes a scaling
solution with $\bar{w}^{\Gamma_{1}}\left(  \bar{A}_{1}\right)  =\frac
{1-\lambda-2\lambda^{2}-6\omega}{3\left(  1+\lambda+2\omega\right)  }$, while
the asymptotic solution is that of the de Sitter universe when $\lambda=-1$
and $\lambda=2$. Furthermore, points $\bar{A}_{2}^{\pm}$ are real for
$\omega<0,~\omega\neq-\frac{3}{2}$ and the points describe stiff fluid
solutions in which $\bar{w}^{\Gamma_{1}}\left(  \bar{A}_{2}^{\pm}\right)  =1$.

\begin{figure}[ptbh]
\centering\includegraphics[width=1\textwidth]{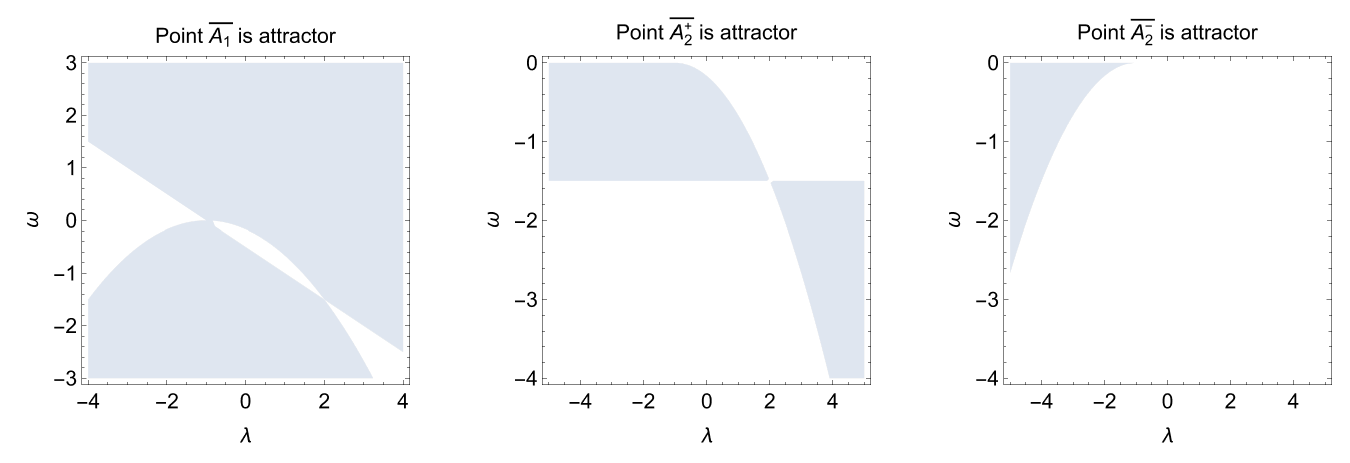}\caption{Region plot in
the space of the free parameters $\left\{  \lambda,\omega\right\}  $ where the
equilibrium points $\bar{A}_{1}$,~$\bar{A}_{2}^{\pm}$ are attractors.}%
\label{ff1}%
\end{figure}

The eigenvalues of the linearized equation (\ref{ds.10}) near the stationary
points are $e\left(  \bar{A}_{1}\right)  =-3+\frac{2+\lambda\left(
1-\lambda\right)  }{1+\lambda+2\omega}$,~$e\left(  \bar{A}_{2}^{\pm}\right)
=\frac{2\left(  3+\lambda\left(  3\pm\sqrt{6\left\vert \omega\right\vert
}\right)  \mp2\sqrt{6\left\vert \omega\right\vert }-6\left\vert \omega
\right\vert \right)  }{3+2\omega}$. Therefore, point $\bar{A}_{1}$ is an
attractor for $\left\{  \lambda\leq-1,\lambda>2,\omega<-\frac{\left(
1+\lambda\right)  ^{2}}{6},\omega>-\frac{1+\lambda}{2}\right\}  $ and
$\left\{  -1<\lambda\leq2,\omega<-\frac{1+\lambda}{2},\omega>-\frac{\left(
1+\lambda\right)  ^{2}}{6}\right\}  $. Similarly, point $\bar{A}_{2}^{+}$ is
an attractor for $\left\{  \lambda<-1,-\frac{3}{2}<\omega<0\right\}  $,
$\left\{  -1<\lambda<2,-\frac{3}{2}<\omega<-\frac{\left(  1+\lambda\right)
^{2}}{6}\right\}  $, $\left\{  \lambda>2,-\frac{\left(  1+\lambda\right)
^{2}}{6}<\omega<-\frac{3}{2}\right\}  $, while point $\bar{A}_{2}^{-}$ is an
attractor for $\left\{  \lambda<-4,-\frac{3}{2}<\omega<0\right\}  $, $\left\{
-4<\lambda<-1,-\frac{\left(  1+\lambda\right)  ^{2}}{6}<\omega<0\right\}  $
and $\left\{  \lambda<-4,-\frac{\left(  1+\lambda\right)  ^{2}}{6}%
<\omega<-\frac{3}{2}\right\}  $. The region plots where the equilibrium points
in the Einstein frame, attractors are presented in Fig. \ref{ff1}.

\begin{table}[tbp] \centering
\caption{Equilibrium points and physical properties for the field equations of the first connection for the Jordan and Einstein frames}%
\begin{tabular}
[c]{ccccc}\hline\hline
\textbf{Point} & \textbf{Existence} & $\mathbf{w}^{\Gamma_{1}}$ &
\textbf{Acceleration?} & \textbf{Attractor?}\\\hline
&  &  &  & \\
\multicolumn{5}{c}{\textbf{Equilibrium points for Connection }$\Gamma_{1}%
$\textbf{ in the Jordan frame}}\\\hline
$A_{1}$ & Always & $-1+\frac{1-\lambda^{2}}{3\omega}$ & Yes & $\omega>0$ ,
$\omega<-\frac{\left(  1+\lambda\right)  ^{2}}{6}$\\
$A_{2}^{+}$ & $\omega<0$ & $1+\sqrt{\frac{8}{3\left\vert \omega\right\vert }}$
& No & $\lambda<-1$ ,$~\left\vert \omega\right\vert <\frac{\left(
1+\lambda\right)  ^{2}}{6}$\\
$A_{2}^{+}$ & $\omega<0$ & $1-\sqrt{\frac{8}{3\left\vert \omega\right\vert }}$
& Yes & $\lambda>1$ ,$~\left\vert \omega\right\vert <\frac{\left(
1+\lambda\right)  ^{2}}{6}$\\
&  &  &  & \\
\multicolumn{5}{c}{\textbf{Equilibrium points for Connection }$\Gamma_{1}%
$\textbf{ in the Einstein frame}}\\\hline
$\bar{A}_{1}$ & Always & $\frac{1-\lambda-2\lambda^{2}-6\omega}{3\left(
1+2\lambda+2\omega\right)  }$ & Yes & Fig. \ref{ff1}\\
$\bar{A}_{2}^{\pm}$ & $\omega<0$ & $1$ & No & Fig. \ref{ff1}\\\hline\hline
\end{tabular}
\label{tabl1}%
\end{table}%

The results of this Section are summarized in Table \ref{tabl1}. We observe
that for this cosmological model, there exists a one-to-one connection between
the stationary points in the two frames. For $\omega>0$, only points $A_{1}$
and $\bar{A}_{1}$ exist. Indeed, every asymptotic solution described by the point
$A_{1}$ reduce to a solution described by the conformal equivalent theory by
point $\bar{A}_{1}$. In general, singular solutions, are transformed into
singular solutions. Except in the case for $\lambda=2$, where the singular
solution at the Jordan frame reads as a de Sitter solution at the Einstein
frame. Moreover, for $\lambda=-1$, the asymptotic solution describes a de
Sitter universe in the two frames. \ \ Furthermore, for $\omega<0$, the
additional points $A_{2}^{\pm}$ and $\bar{A}_{2}^{\pm}$ exist. In the Einstein
frame at these points, the asymptotic solutions describe only stiff fluid
components, while in the Jordan frame, other fluid components can be described.

\section{Phase-space analysis for connection $\Gamma_{2}$}

\label{sec5}

We proceed our analysis with the field equations which correspond to the
selection for the connection $\Gamma_{2}$. Indeed, from the point-like
Lagrangian (\ref{rs.02}) we determine the field equations \cite{anbb}%
\begin{align}
3H^{2}+\frac{\omega}{2}\dot{\phi}^{2}+\frac{3}{2}\dot{\phi}\dot{\psi}%
+e^{-\phi}V\left(  \phi\right)   &  =0,\label{jo.01}\\
2\dot{H}+3H^{2}+2H\dot{\phi}-\frac{\omega}{2}\dot{\phi}^{2}-\frac{3}{2}%
\dot{\phi}\dot{\psi}+e^{-\phi}V\left(  \phi\right)   &  =0,\label{jo.02}\\
3\ddot{\psi}+2\omega\ddot{\phi}+H\left(  6\omega\dot{\phi}+9\dot{\psi}\right)
-6H^{2}+\omega\dot{\phi}^{2}+e^{-\phi}V_{,\phi} &  =0,\label{jo.03}\\
\ddot{\phi}+\dot{\phi}^{2}+3H\dot{\phi} &  =0,\label{jo.04}%
\end{align}
where the effective fluid components are%
\begin{align}
\rho_{eff}\left(  \Gamma_{2}\right)   &  =-\left(  \frac{\omega}{2}\dot{\phi
}^{2}+\frac{3}{2}\dot{\phi}\dot{\psi}+e^{-\phi}V\left(  \phi\right)  \right)
\\
p_{eff}\left(  \Gamma_{2}\right)   &  =-\left(  \frac{\omega}{2}\dot{\phi}%
^{2}+\frac{3}{2}\dot{\phi}\dot{\psi}-e^{-\phi}V\left(  \phi\right)  \right)
+2H\dot{\phi}.
\end{align}
and $N\left(  t\right)  =1$.

We work in the dimensionless variables%
\[
\tau,~~x,~~y~\text{and }z=\sqrt{\frac{3}{2}}\frac{\dot{\psi}}{H},
\]
where the field equations (\ref{jo.01})-(\ref{jo.04}) are expressed as%
\begin{align}
\frac{dx}{d\tau}  &  =-\frac{3}{2}x\left(  1-y+x\left(  \omega x+z\right)
\right), \label{jo.05}\\
\frac{dy}{d\tau}  &  =y\left(  3\left(  1+y\right)  +x\left(  \sqrt{6}\left(
1+\lambda\right)  -3\left(  \omega x+z\right)  \right)  \right)
,\label{jo.06}\\
\frac{dz}{d\tau}  &  =\frac{1}{2}\left(  3\left(  y-1\right)  z+\left(
xz+\omega x^{2}\right)  \left(  2\sqrt{6}-3z\right)  -2\sqrt{6}\left(  \lambda
y-1\right)  \right),  \label{jo.07}%
\end{align}
and constraint equation%
\begin{equation}
1+y+x\left(  \omega x+z\right)  =0. \label{jo.08}%
\end{equation}

Finally, the equation of state parameter is expressed as
\begin{equation}
w^{\Gamma_{2}}\left(  x,y,z\right)  =y-\frac{1}{3}x\left(  3\left(  \omega
x+z\right)  -2\sqrt{6}\right). 
\end{equation}

By applying the constraint equation (\ref{jo.08}) the dimension of the
dynamical system is reduced by one, and the stationary points are defined in
the plane $B=\left(  x\left(  B\right), z\left(  B\right)  \right)  $.

They are%
\[
B_{1}=\left(  x_{1},-\frac{1}{x_{1}}-\omega x_{1}\right), ~~B_{2}=\left(
0,\sqrt{\frac{2}{3}}\left(  1+\lambda\right)  \right), 
\]
where $x_{1}$ in $B_{1}$ is an arbitrary constant. Specifically $B_{1}$
describes a family of points with the equation of state parameter $w^{\Gamma_{2}%
}\left(  B_{1}\right)  =1+\sqrt{\frac{8}{3}}x_{1}$. Moreover, point $B_{2}$
describes the de Sitter universe with $~w^{\Gamma_{2}}\left(  B_{2}\right)
=-1$.

The eigenvalues of the two-dimensional linearized system around the stationary
points are $e_{1}\left(  B_{1}\right)  =0,~~e_{2}\left(  B_{1}\right)
=\sqrt{6}\left(  \sqrt{6}+\left(  1+\lambda\right)  x_{1}\right)  $ and
$e_{1}\left(  B_{2}\right)  =-3,~~e_{2}\left(  B_{2}\right)  =-3$. As a
result, point $B_{2}$ is always an attractor, while for the family of points
$B_{1}~$because one of the eigenvalues has zero real part, the Center Manifold
Theorem (CMT) should be applied. From the latter, we will be able to show if
there exists any stable submanifold when $e_{2}\left(  B_{1}\right)  <0$.

In order to calculate the CMT, we perform the change of variable $z=-\frac
{1}{x}-\omega x+\tilde{z}$, such that the dynamical system reduced to the
following form
\begin{align}
\frac{dx}{d\tau}  &  =-3x^{2}\tilde{z},\\
\frac{d\tilde{z}}{d\tau}  &  =\left(  6+x\left(  \sqrt{6}\left(
1+\lambda\right)  -3z\right)  \right)  \tilde{z},
\end{align}
where in the new variables points $B_{1}$ have coordinates $\tilde{z}=0$. In
order to determine the stable manifold, we assume that $\tilde{z}=h\left(
x\right)  $, where we end with the equation%
\begin{equation}
h\left(  6+\sqrt{6}x\left(  1+\lambda\right)  -3xh\left(  x\right)
+3x^{2}\frac{dh\left(  x\right)  }{dx}\right)  =0,
\end{equation}
with solutions $h\left(  x\right)  =0$, and $h\left(  x\right)  =\frac{1}%
{3}\left(  \frac{3}{x}+\sqrt{6}\left(  1+\lambda\right)  \right)  +h_{1}x$.

In order $h\left(  x\right)  $ to describe a stable submanifold it should
hold, $h\left(  x_{1}\right)  =0$ and $\frac{dh}{dx}|_{x=x_{1}}=0$.
Consequently, the unique stable submanifold is the surface of points with
$h\left(  x\right)  =0$. That means, that if the initial conditions belong
to the family of points $B_{1}$, for $e_{2}\left(  B_{1}\right)  <0$, the
trajectory solutions will stay on the surface defined by \ points $B_{1}$.

Nevertheless, variables $\left\{  x,z\right\}  $ are not constrained which
means that they can take values at infinity. Thus, we should determine the
existence of stationary points at the infinity regime. In order to perform
such analysis we introduce the Poincare variables%
\[
x=\frac{X}{\sqrt{1-X^{2}-Z^{2}}},~~z=\frac{Z}{\sqrt{1-X^{2}-Z^{2}}}%
,~~dT=\sqrt{1-X^{2}-Z^{2}}d\tau,
\]
and we write the two-dimensional dynamical system in the form
\[
\frac{dX}{dT}=f_{1}\left(  X,Z\right), ~~\frac{dZ}{dT}=f_{2}\left(
X,Z\right). 
\]

Infinity is reached when $1-X^{2}-Z^{2}=0$, thus, the admitted equilibrium
points $B^{\inf}=\left(  X\left(  B^{\inf}\right), Z\left(  B^{\inf}\right)
\right)  $ at the infinity are%
\[
B_{1\pm}^{\inf}=\left(  0,\pm1\right), ~~B_{2\pm}^{\inf}=\left(  \pm
\sqrt{\frac{1}{1+\omega^{2}}},\frac{\omega}{\sqrt{1+\omega^{2}}}\right)
,~~B_{3\pm}^{\inf}=\left(  \pm\sqrt{\frac{1}{1+\omega^{2}}},-\frac{\omega
}{\sqrt{1+\omega^{2}}}\right)  \text{.}%
\]

We derive that the stationary points $B_{1\pm}^{\inf}$ describe de Sitter
universes, that is, $w^{\Gamma_{2}}\left(  B_{1\pm}^{\inf}\right)  =-1$, while
points $B_{2\pm}^{\inf}$ and $~B_{3\pm}^{\inf}$ correspond to Big Rip
singularities, that is, $w^{\Gamma_{2}}\left(  ~B_{2\pm}^{\inf}\right)
=-\infty$ and $w^{\Gamma_{2}}\left(  ~B_{3\pm}^{\inf}\right)  =-\infty$. 
As far as stability is concerned, it follows that all the stationary points
at the infinity describe unstable solutions.

\begin{figure}[ptbh]
\centering\includegraphics[width=1\textwidth]{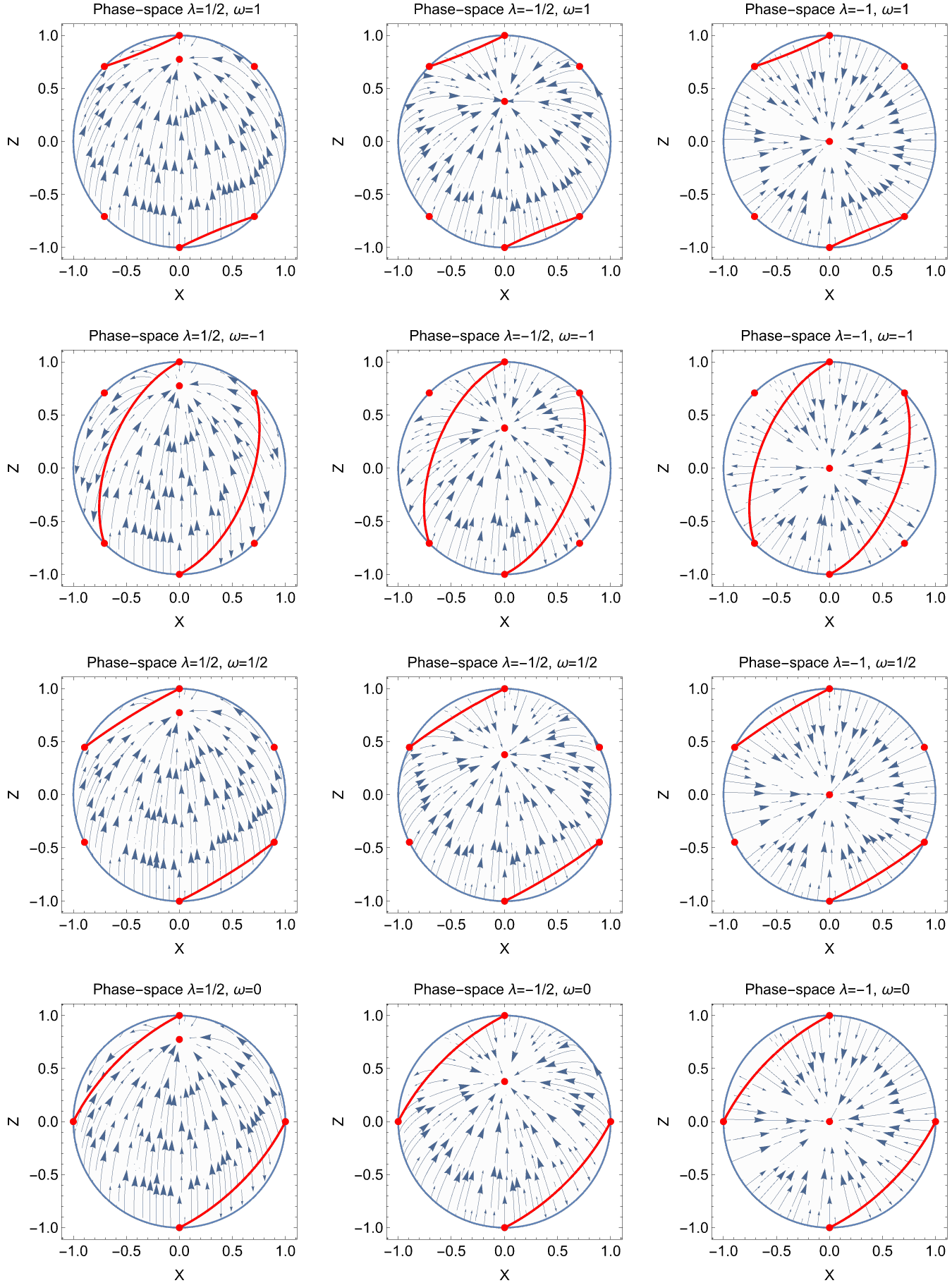}\caption{Phase-space
portrait for the field equations for the connection $\Gamma_{2}$ in the Jordan
frame in the Poincare variables. With red are marked the stationary points,
and red lines correspond to the family of points $B_{1}$. We observe that
$B_{2}$ is the unique attractor of the cosmological model. We observe that in
order for the cosmological evolution not to suffer from a Big Rip singularity in
the future, we should start from initial conditions inside the region bounded
by the family of points $B_{1}$. }%
\label{ff2}%
\end{figure}

In Fig. \ref{ff2} we present phase-space portraits for this dynamical system
for different values of the free parameters $\omega$ and $\lambda$. We observe
that in order the cosmological evolution not to suffer from a Big Rip
singularity in the future, we should start from the initial conditions inside the
region bounded by the family of points $B_{1}$.

\subsection{Conformal equivalent theory}

We proceed with the analysis of the dynamics for the conformal equivalent
theory described by the Lagrangian function (\ref{rs.06}).

For this cosmological model, the cosmological field equations are \cite{anbb}
\begin{align}
3\bar{H}^{2}-3\bar{H}\dot{\phi}+\frac{\bar{\omega}}{2}\dot{\phi}^{2}+\frac
{3}{2}\dot{\phi}\dot{\psi}+\bar{V}\left(  \phi\right)   &  =0,\label{ef.01}\\
2\left(  \bar{H}\right)  ^{\cdot}+3\bar{H}^{2}-\ddot{\phi}-\frac{\bar{\omega}%
}{2}\dot{\phi}^{2}-\frac{3}{2}\dot{\phi}\dot{\psi}+\bar{V}\left(  \phi\right)
&  =0,\label{ef.02}\\
2\left(  \bar{H}\right)  ^{\cdot}+6\bar{H}^{2}-\frac{2}{3}\bar{V}_{,\phi
}\left(  \phi\right)  -\left(  \ddot{\psi}+3\bar{H}\dot{\psi}\right)
-2\bar{\omega}\dot{\phi}H &  =0,\label{ef.03}\\
\ddot{\phi}+3\bar{H}\dot{\phi} &  =0,\label{ef.04}%
\end{align}
with effective fluid components%
\begin{align}
\bar{\rho}_{eff}\left(  \Gamma_{2}\right)   &  =-\left(  \frac{\bar{\omega}%
}{2}\dot{\phi}^{2}+\frac{3}{2}\dot{\phi}\dot{\psi}+\bar{V}\left(  \phi\right)
\right)  +3\bar{H}\dot{\phi},\\
\bar{p}_{eff}\left(  \Gamma_{2}\right)   &  =-\left(  \frac{\bar{\omega}}%
{2}\dot{\phi}^{2}+\frac{3}{2}\dot{\phi}\dot{\psi}-\bar{V}\left(  \phi\right)
\right)  -\ddot{\phi}.
\end{align}
and $\bar{N}\left(  t\right)  =1$.

We work in the dimensionless variables
\[
\bar{\tau},~~\bar{x},~~\bar{y},~~\bar{z}=\sqrt{\frac{3}{2}}\frac{\dot{\psi}%
}{\bar{H}},
\]
where now the field equations are expressed as
\begin{align}
\frac{d\bar{x}}{d\bar{\tau}} &  =-\frac{3}{2}\bar{x}\left(  1-\bar{y}+\bar
{x}\left(  \bar{\omega}\bar{x}+\bar{z}-\sqrt{6}\right)  \right), \\
\frac{d\bar{y}}{d\bar{\tau}} &  =\bar{y}\left(  3\left(  1+\bar{y}\right)
+x\left(  \sqrt{6}\left(  1+\lambda\right)  -3\left(  \bar{\omega}\bar
{x}+z\right)  \right)  \right), \\
\frac{d\bar{z}}{d\bar{\tau}} &  =\frac{1}{2}\left(  \sqrt{6}\left(  3+\bar
{y}\left(  1-2\lambda\right)  \right)  +3\bar{\omega}\bar{x}^{2}\left(
\sqrt{6}-\bar{z}\right)  +3\bar{z}\left(  \bar{y}-1\right)  +3\bar{x}\left(
2\sqrt{6}-\bar{z}\right)  \bar{z}-18x\right), 
\end{align}
and%
\begin{equation}
1+\bar{y}+\bar{x}\left(  \bar{\omega}\bar{x}+\bar{z}-\sqrt{6}\right)  =0,
\end{equation}
with equation of state parameter%
\[
\bar{w}^{\Gamma_{2}}\left(  \bar{x},\bar{y},\bar{z}\right)  =\bar{y}-\bar
{x}\left(  \bar{\omega}\bar{x}+\bar{z}-\sqrt{6}\right)
\]

The stationary points of the latter system are defined in the two-dimensional
manifold $\bar{B}=\left(  \bar{x}\left(  \bar{B}\right), \bar{z}\left(
\bar{B}\right)  \right)  $; they are
\begin{equation}
\bar{B}_{1}=\left(  x_{1},\sqrt{6}-\bar{\omega}x_{1}-\frac{1}{x_{1}}\right)
,~~\bar{B}_{2}=\left(  0,\sqrt{\frac{2}{3}}\left(  1+\lambda\right)  \right)
.
\end{equation}

Points $\bar{B}_{1}$ describe a family of points which exist for $x_{1}\neq0$.
The asymptotic solutions at the points correspond to universes dominated by a
stiff fluid, i.e. $\bar{w}^{\Gamma_{2}}\left(  \bar{B}_{1}\right)  =1$.
Moreover, Point $\bar{B}_{2}$ describes a de Sitter solution, $\bar{w}%
^{\Gamma_{2}}\left(  \bar{B}_{2}\right)  =-1$ which is a future attractor for
the dynamical system; since the eigenvalues of the linearized system are
$e_{1}\left(  \bar{B}_{2}\right)  =-3,~~e_{2}\left(  \bar{B}_{2}\right)  =-3$.
As far as the stability properties of points $\bar{B}_{1}$ are concerned, we
determine the two eigenvalues $e_{1}\left(  \bar{B}_{1}\right)  =0,~~e_{2}%
\left(  \bar{B}_{1}\right)  =6+\sqrt{6}\left(  \lambda-2\right)  x_{1}$.
Because $e_{1}\left(  \bar{B}_{1}\right)  $ is zero, we apply the CMT as
before and we found that the stationary points do not describe stable
solutions, except if the initial conditions are that defined on the family
of points $\bar{B}_{1}$. \

We remark that at the finite regime, there exists a one-to-one correspondence
between the equilibrium points, and their asymptotic solutions, for the two
conformal equivalent theories defined in the Jordan and the Einstein frames.
We proceed with the analysis of the asymptotics at the infinite regime.

We define the Poincare variables%
\[
\bar{x}=\frac{\bar{X}}{\sqrt{1-\bar{X}^{2}-\bar{Z}^{2}}},~~\bar{z}=\frac
{\bar{Z}}{\sqrt{1-\bar{X}^{2}-\bar{Z}^{2}}},~~d\bar{T}=\sqrt{1-\bar{X}%
^{2}-\bar{Z}^{2}}d\bar{\tau}.
\]
Hence, the dynamical system can be written in the following form
\[
\frac{d\bar{X}}{d\bar{T}}=\bar{f}_{1}\left(  \bar{X},\bar{Z}\right)
,~~\frac{d\bar{Z}}{d\bar{T}}=\bar{f}_{2}\left(  \bar{X},\bar{Z}\right). 
\]

At infinity, the stationary points $\bar{B}^{\inf}=\left(  \bar{X}\left(
\bar{B}^{\inf}\right), \bar{Z}\left(  \bar{B}^{\inf}\right)  \right)  $ are
\[
\bar{B}_{1\pm}^{\inf}=\left(  0,\pm1\right), ~~\bar{B}_{2\pm}^{\inf}=\left(
\pm\sqrt{\frac{1}{1+\bar{\omega}^{2}}},\frac{\bar{\omega}}{\sqrt{1+\bar
{\omega}^{2}}}\right), ~~\bar{B}_{3\pm}^{\inf}=\left(  \pm\sqrt{\frac
{1}{1+\bar{\omega}^{2}}},-\frac{\omega}{\sqrt{1+\bar{\omega}^{2}}}\right). 
\]
Similar to the conformal equivalent theory defined in the in Jordan frame,
points $\bar{B}_{1\pm}^{\inf}$ describe de Sitter solutions, $\bar{w}%
^{\Gamma_{2}}\left(  \bar{B}_{1\pm}^{\inf}\right)  $ while points $\bar
{B}_{2\pm}^{\inf}$ and $\bar{B}_{3\pm}^{\inf}$ correspond to Big Rip
singularities, i.e. $\bar{w}^{\Gamma_{2}}\left(  ~\bar{B}_{2\pm}^{\inf
}\right)  =-\infty$ and $\bar{w}^{\Gamma_{2}}\left(  ~\bar{B}_{3\pm}^{\inf
}\right)  =-\infty$. We omit the presentation of the stability analysis, but
we conclude that all the stationary points at the infinity describe unstable solutions.

Thus, the unique attractor for this model is the de Sitter universe described
by point $B_{2}$. Additionally, we remark that there exists an one-to-one
corresponds to the equilibrium points between the Jordan and the Einstein
frames. The only physical solution which does not remain invariant is that
described by points $B_{1}$. Indeed the conformal equivalent points $\bar
{B}_{1}$ describe only stiff fluid solutions, while the solutions at the
family of point $B_{1}$ can describe accelerated universes.

The results of this Section are summarized in Table \ref{tabl2}, where the
physical properties of the stationary points can be compared between the two frames.%

\begin{table}[tbp] \centering
\caption{Equilibrium points and physical properties for the field equations of the second connection for the Jordan and Einstein frames}%
\begin{tabular}
[c]{ccccc}\hline\hline
\textbf{Point} & \textbf{Existence} & $\mathbf{w}^{\Gamma_{2}}$ &
\textbf{Acceleration?} & \textbf{Attractor?}\\\hline
&  &  &  & \\
\multicolumn{5}{c}{\textbf{Equilibrium points for Connection }$\Gamma_{2}%
$\textbf{ in the Jordan frame}}\\\hline
$B_{1}$ & $x\neq0$ & $1+\sqrt{\frac{8}{3}}x_{1}$ & Yes & No\\
$B_{2}$ & Always & $-1$ & Always & Always\\
$B_{1\pm}^{\inf}$ & Always & $-1$ & Always & No\\
$B_{2\pm}^{\inf}$ & Always & Big Rip & Always & No\\
$B_{3\pm}^{\inf}$ & Always & Big Rip & Always & No\\
&  &  &  & \\
\multicolumn{5}{c}{\textbf{Equilibrium points for Connection }$\Gamma_{2}%
$\textbf{ in the Einstein frame}}\\\hline
$\bar{B}_{1}$ & $\bar{x}\neq0$ & $1$ & No & No\\
$\bar{B}_{2}$ & Always & $-1$ & Always & Always\\
$\bar{B}_{1\pm}^{\inf}$ & Always & $-1$ & Always & No\\
$\bar{B}_{2\pm}^{\inf}$ & Always & Big Rip & Always & No\\
$\bar{B}_{3\pm}^{\inf}$ & Always & Big Rip & Always & No\\\hline\hline
\end{tabular}
\label{tabl2}%
\end{table}%

\section{Phase-space analysis for connection $\Gamma_{3}$}

\label{sec6}

Finally, for the third connection and Lagrangian function (\ref{rs.03}) we
derive the field equations
\begin{align}
3H^{2}+\frac{\omega}{2}\dot{\phi}^{2}-\frac{3}{2a^{2}}\frac{\dot{\phi}}%
{\dot{\Psi}}+e^{-\phi}V\left(  \phi\right)   &  =0,\\
2\dot{H}+3H^{2}+2H\dot{\phi}-\frac{\omega}{2}\dot{\phi}^{2}-\frac{1}{2a^{2}%
}\frac{\dot{\phi}}{\dot{\Psi}}+e^{-\phi}V\left(  \phi\right)   &  =0,\\
\frac{3}{2a^{2}}\left(  \ddot{\Psi}-H\dot{\Psi}\right)  +\dot{\Psi}^{2}\left(
3\left(  H^{2}-\omega H\dot{\phi}\right)  -\frac{\omega}{2}\left(  \dot{\phi
}^{2}+2\ddot{\phi}\right)  \right)  -e^{-\phi}V_{,\phi}\left(  \phi\right)
\dot{\Psi}^{2} &  =0,\\
\dot{\Psi}\left(  \dot{\phi}\left(  H+\dot{\phi}\right)  +\ddot{\phi}\right)
-2\dot{\phi}\ddot{\Psi} &  =0,
\end{align}
and the fluid components are expressed as follows
\begin{align}
\rho_{eff}\left(  \Gamma_{3}\right)   &  =-\left(  \frac{\omega}{2}\dot{\phi
}^{2}-\frac{3}{2a^{2}}\frac{\dot{\phi}}{\dot{\Psi}}+e^{-\phi}V\left(
\phi\right)  \right), \\
p_{eff}\left(  \Gamma_{3}\right)   &  =-\left(  \frac{\omega}{2}\dot{\phi}%
^{2}+\frac{1}{2a^{2}}\frac{\dot{\phi}}{\dot{\Psi}}-e^{-\phi}V\left(
\phi\right)  \right)  +2H\dot{\phi},
\end{align}
with $N\left(  t\right)  =1$.

In the dimensionless variables we work in the dimensionless variables
\[
\tau,~~x,~~y,~~\xi=\sqrt{\frac{2}{3}}\frac{1}{a^{2}\dot{\Psi}H},~
\]
the field equations become
\begin{align}
\frac{dx}{d\tau}  &  =\frac{x}{4}\left(  10-6\left(  \omega x^{2}-y\right)
-3x\xi+\frac{16\left(  \omega x\left(  \sqrt{6}x-8\right)  +\sqrt{6}\left(
1-\lambda y\right)  \right)  }{8\omega x-3\xi}\right), \\
\frac{dy}{d\tau}  &  =\frac{y}{2}\left(  6\left(  1+y\right)  +x\left(
2\sqrt{6}\left(  1+\lambda\right)  -3\left(  2\omega x+\xi\right)  \right)
\right), \\
\frac{d\xi}{d\tau}  &  =\frac{\xi\left(  8\omega x\left(  2+3x\left(  \sqrt
{6}-2\omega x\right)  +6y\right)  -8\sqrt{6}\left(  1-\lambda y\right)
-6\xi\left(  2\sqrt{6}x+\omega x^{2}+3\left(  y-1\right)  \right)  +9x\xi
^{2}\right)  }{4\left(  8\omega x-3\xi\right)  },
\end{align}
and%
\begin{equation}
1+\omega x^{2}+y-\frac{3}{2}x\xi=0. \label{kk.04}%
\end{equation}

Thus, the equation of state parameter reads%
\begin{equation}
w^{\Gamma_{3}}\left(  x,y,\xi\right)  =y-\omega x^{2}+\frac{x}{6}\left(
4\sqrt{6}-3\xi\right). 
\end{equation}

With the use of the algebraic equation (\ref{kk.04}) the dynamical system is
reduced to a two-dimensional system, where the stationary points $C=\left(
x\left(  C\right), \xi\left(  C\right)  \right)  $ in the finite regime are
\[
C_{1}=\left(  \sqrt{\frac{2}{3}}\frac{5}{1-3\lambda},\frac{\sqrt{6}\left(
10\omega-1+\lambda\left(  2+3\lambda\right)  \right)  }{3\left(
1-3\lambda\right)  }\right), ~C_{2}=\left(  0,\sqrt{\frac{8}{27}}\left(
1+\lambda\right)  \right), 
\]%
\[
C_{3}^{\pm}=\left(  \frac{1\pm\sqrt{1-2\omega}}{\sqrt{6}\omega},\sqrt{\frac
{8}{27}}\left(  2\mp\sqrt{1-2\omega}\right)  \right), 
\]%
\[
C_{4}=\left(  \frac{1+\lambda}{\sqrt{6}\omega},0\right), ~~C_{5}^{\pm
}=\left(  \pm\frac{1}{\sqrt{-\omega}},0\right). 
\]

Point $C_{2}$ exist always, however, for the rest of the points the existence
conditions are, for point $C_{1}$, $\lambda\neq\frac{1}{3}$; for points
$C_{3}^{\pm}$,~$\omega\neq0$ and $\omega<\frac{1}{2}$; point $C_{4}$ exists
for $\omega\neq0$. Finally, points $C_{5}^{\pm}$ are real when $\omega<0$. The
equation of state parameter for the effective fluid at the asymptotic
solutions at the equilibrium points are $w^{\Gamma_{3}}\left(  C_{1}\right)
=\frac{\lambda-7}{3\left(  3\lambda-1\right)  },~w^{\Gamma_{3}}\left(
C_{2}\right)  =-1$, $w^{\Gamma_{3}}\left(  C_{3}^{\pm}\right)  =\frac{1}%
{9}+\frac{2\left(  1\pm\sqrt{1-2\omega}\right)  }{9\omega}$, $w^{\Gamma_{3}%
}\left(  C_{4}\right)  =-1+\frac{1-2\lambda}{3\omega}$ and $w^{\Gamma_{3}%
}\left(  C_{5}^{\pm}\right)  =1\pm\sqrt{-\frac{8}{3\omega}}$.

Point $C_{1}$ describes a scaling solution, where acceleration is occurred for
$\,\frac{1}{3}<\lambda<2$, and for $\lambda=1$ the de Sitter universe is
recovered. Furthermore, $C_{2}$ corresponds to the de Sitter point, similar to
point $B_{2}$. Points $C_{3}^{\pm}$ describe scaling solutions, accelerated is
occurred for $-\frac{3}{2}<\omega<0$. Last but not least, points $C_{4}$ and
$C_{5}^{\pm}$ have the same physical properties with points $A_{1}$ and
$A_{2}^{\pm}$ respectively.

\begin{figure}[ptbh]
\centering\includegraphics[width=1\textwidth]{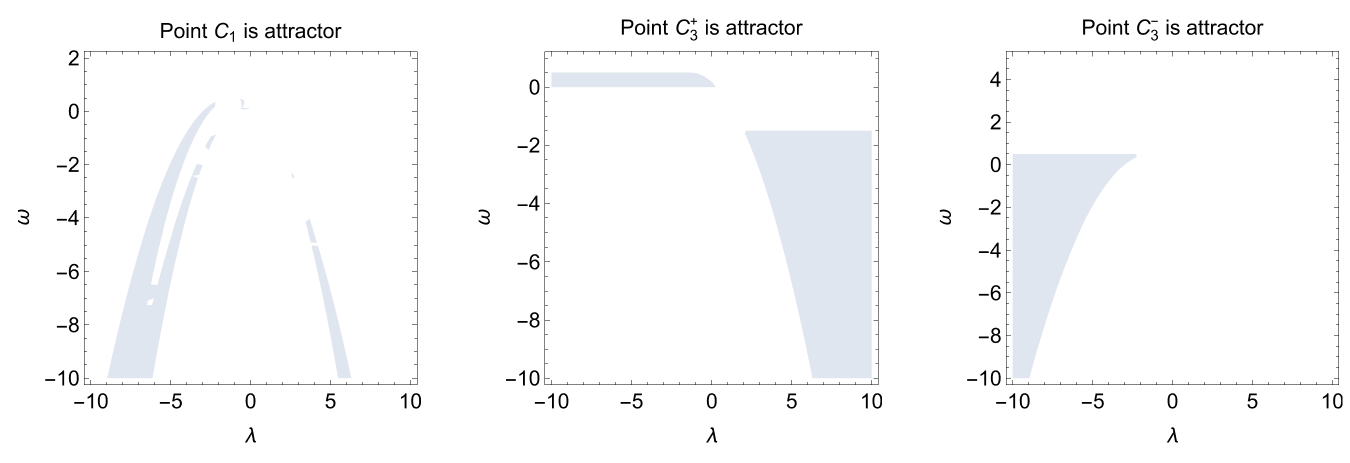}\caption{Region plot in
the space of the free parameters $\left\{  \lambda,\omega\right\}  $ where the
equilibrium points $C_{1}$ and~$C_{3}^{\pm}$ are attractors.}%
\label{ff3}%
\end{figure}

As far as the stability properties of the stationary points are concerned, in
Fig. \ref{ff3} we present the regions in the space of the free variables
$\left\{  \lambda,\omega\right\}  $ in which points $C_{1}$ and $C_{3}^{\pm}$
are attractors. For point, $C_{2}$, the eigenvalues of the linearized system
have always negative real parts which means that the de Sitter solution is a
future attractor. Furthermore, for point $C_{4}$ we calculate the eigenvalues
$e_{1}\left(  C_{4}\right)  =-3-\frac{\left(  1+\lambda\right)  ^{2}}{2\omega
}$ and $e_{2}\left(  C_{4}\right)  =\frac{1-\lambda\left(  2+3\lambda\right)
-10\omega}{4\omega}$, from where it follows that the equilibrium point $C_{4}$
is an attractor when $\left\{  \lambda\leq-1,\lambda>2,\omega<\frac
{1-2\lambda-3\lambda^{2}}{10},\omega>0\right\}  $, $\left\{  -1<\lambda
\leq-\frac{1}{3},\frac{1-2\lambda-3\lambda^{2}}{10}<\omega,~~\omega
<-\frac{\left(  1+\lambda\right)  ^{2}}{6}\right\}  $, $\left\{  \frac{1}%
{3}<\lambda\leq2,\omega>0,\omega<-\frac{\left(  1+\lambda\right)  ^{2}}%
{6}\right\}  $. Finally, for points $C_{5}^{\pm}$ the eigenvalues are
$e_{1}\left(  C_{5}^{\pm}\right)  =2\pm\sqrt{\frac{6}{-\omega}}\,\ $,
$e_{2}\left(  C_{5}^{\pm}\right)  =\sqrt{6}\left(  \sqrt{6}\pm\frac{\left(
1+\lambda\right)  }{\sqrt{-\omega}}\right)  $, from where we conclude that the
solution at $C_{5}^{+}$ is always unstable and $C_{5}^{-}$ is an attractor
when $\left\{  -1<\lambda\leq2,\left\vert \omega\right\vert <\frac{\left(
1+\lambda\right)  ^{2}}{2}\right\}  $ and $\left\{  \lambda>2,-\frac{3}%
{2}<\omega<0\right\}  $.

We continue with the analysis of the dynamics at the infinity. We make use of
the Poincar\'{e} variables%
\[
x=\frac{X}{\sqrt{1-X^{2}-\Xi^{2}}},~\xi=\frac{\Xi}{\sqrt{1-X^{2}-\Xi^{2}}%
},
\]
and the time derivative 
\[~\frac{d f}{dT}=\sqrt{1-X^{2}-\Xi^{2}}\frac{d f}{d\tau},\]
to obtain the dynamical system is written in the form
\[
\frac{dX}{dT}=g_{1}\left(  X,\Xi\right), ~~\frac{d\Xi}{dT}=g_{2}\left(
X,\Xi\right).
\]

The stationary points of the latter dynamical system at infinity are%
\[
C_{1\pm}^{\inf}=\left(  0,\pm1\right), ~~C_{2\pm}^{\inf}=\left(  \pm
\sqrt{\frac{9}{9+4\omega^{2}}},\frac{2\omega}{\sqrt{9+4\omega^{2}}}\right)
,~~C_{3\pm}^{\inf}=\left(  \pm\sqrt{\frac{9}{9+4\omega^{2}}},-\frac{2\omega
}{\sqrt{9+4\omega^{2}}}\right). 
\]
Stationary points $C_{1\pm}^{\inf}$ describe de Sitter solutions and
$~C_{2\pm}^{\inf}$, $C_{3\pm}^{\inf}$ correspond to Big Rip singularities. It
is straightforward to show that the equilibrium points at infinity do not
describe any stable solution.

\subsection{Conformal equivalent theory}

We proceed with the investigation of the equilibrium points for the conformal
equivalent theory.\ Indeed, from Lagrangian (\ref{rs.07}) we derive the field
equations%
\begin{align}
3\bar{H}^{2}-3\bar{H}\dot{\phi}+\frac{\bar{\omega}}{2}\dot{\phi}-\frac
{3}{2\alpha^{2}}\frac{\dot{\phi}}{\dot{\Psi}}+\bar{V}\left(  \phi\right)   &
=0,\\
2\left(  \bar{H}\right)  ^{\cdot}+3\bar{H}^{2}-\frac{\bar{\omega}}{2}\dot
{\phi}^{2}+V\left(  \phi\right)  -\ddot{\phi}-\frac{1}{2\alpha^{2}}\frac
{\dot{\phi}}{\dot{\Psi}} &  =0,\\
\frac{3}{2\alpha^{2}}\left(  \bar{H}\dot{\Psi}-\ddot{\Psi}\right)  -3\left(
\left(  \bar{H}\right)  ^{\cdot}+3\bar{H}^{2}\right)  +\ddot{\omega}\left(
\ddot{\phi}+3\bar{H}\dot{\phi}\right)  +\bar{V}_{,\phi} &  =0,\\
\dot{\Psi}\left(  \ddot{\phi}+\bar{H}\dot{\phi}\right)  -2\dot{\phi}\ddot
{\Psi} &  =0,
\end{align}
for $\bar{N}\left(  t\right)  =1$.

From the latter set of field equations, we define the components%
\begin{align}
\bar{\rho}_{eff}\left(  \Gamma_{3}\right)   &  =-\left(  \frac{\omega}{2}%
\dot{\phi}^{2}-\frac{3}{2\alpha^{2}}\frac{\dot{\phi}}{\dot{\Psi}}+\bar
{V}\left(  \phi\right)  \right)  +3\bar{H}\dot{\phi},\\
\bar{p}_{eff}\left(  \Gamma_{3}\right)   &  =-\left(  \frac{\omega}{2}%
\dot{\phi}^{2}+\frac{1}{2\alpha^{2}}\frac{\dot{\phi}}{\dot{\Psi}}-\bar
{V}\left(  \phi\right)  \right)  -\ddot{\phi}.
\end{align}

In terms of the dimensionless variables%
\[
\bar{\tau},~~\bar{x},~~\bar{y}~\text{and }\bar{\xi}=\xi\text{,}%
\]
we write the following dynamical system%
\begin{align}
\frac{2\left(  8\left(  2\bar{\omega}-3\right)  x-6\bar{\xi}\right)  }{\bar
{x}}\frac{d\bar{x}}{d\bar{\tau}}  &  =3\bar{x}\left(  16\left(  \bar{y}\left(
\lambda-2+\bar{\omega}\right)  -\left(  \bar{\omega}+3\right)  \right)
+3\bar{\xi}\left(  2\sqrt{6}+\bar{\xi}\right)  \right)  +8\bar{\omega}\left(
1-\bar{x}\right) \nonumber\\
&  -48\bar{\omega}^{2}\bar{x}^{3}+8\sqrt{6}\left(  3+\bar{y}\left(
1-2\lambda\right)  \right)  -6\bar{\xi}\left(  5+3\bar{y}\right)
+6\bar{\omega}\bar{x}^{2}\left(  12\sqrt{6}-\xi\right), \\
\frac{\left(  8\left(  2\bar{\omega}-3\right)  x-6\bar{\xi}\right)  }{\bar{y}%
}\frac{d\bar{y}}{d\bar{\tau}}  &  =3\bar{x}\left(  16\left(  \bar{y}\left(
\lambda-2+\bar{\omega}\right)  +\left(  \bar{\omega}-3\right)  \right)
+\bar{\xi}\left(  3\bar{\xi}-2\sqrt{6}\left(  \lambda-3\right)  \right)
\right) \\
&  -48\bar{\omega}^{2}\bar{x}^{3}-18\bar{\xi}\left(  1+\bar{y}\right)
+2\bar{x}^{2}\left(  12\sqrt{6}\left(  2-\lambda+3\bar{\omega}\left(
1+\lambda\right)  \right)  -3\bar{\omega}\xi\right), \\
\frac{2\left(  8\left(  2\bar{\omega}-3\right)  x-6\bar{\xi}\right)  }%
{\bar{\xi}}\frac{d\bar{\xi}}{d\bar{\tau}}  &  =\bar{x}\left(  4\left(
4\bar{\omega}-6\right)  +48\left(  \bar{\omega}+\lambda-2\right)  \bar
{y}+9\bar{\xi}^{2}\right)  -48\bar{\omega}^{2}\bar{x}^{3}+6\bar{\omega}\bar
{x}^{2}\left(  6\sqrt{6}-\bar{\xi}\right) \\
&  +2\left(  \bar{y}\left(  2\sqrt{6}\left(  2\lambda-1\right)  -9\bar{\xi
}\right)  +9\bar{\xi}-6\sqrt{6}\right), 
\end{align}
and constraint%
\begin{equation}
1+\bar{\omega}\bar{x}^{2}+\bar{y}-\frac{1}{2}\bar{x}\left(  2\sqrt{6}%
+3\bar{\xi}\right)  =0.
\end{equation}
Moreover, we calculate the effective equation of state parameter%
\[
\bar{w}^{\Gamma_{3}}\left(  \bar{x},\bar{y},\bar{\xi}\right)  =\frac
{2\bar{\omega}\bar{x}^{2}\left(  8\sqrt{6}-\bar{\xi}\right)  -6\bar{y}\bar
{\xi}+\bar{x}\left(  16\left(  \bar{\omega}+\lambda-2\right)  \bar{y}%
+2\sqrt{6}\bar{\xi}+3\bar{\xi}-24\right)  -16\bar{\omega}^{2}x^{3}}{\left(
8\left(  2\bar{\omega}-3\right)  x-6\bar{\xi}\right)  }.
\]

The stationary points $\bar{C}=\left(  \bar{x}\left(  \bar{C}\right)
,\bar{\xi}\left(  \bar{C}\right)  \right)  $ of the latter dynamical system
have the following coordinates%
\[
\bar{C}_{1}=\left(  \sqrt{\frac{2}{3}}\frac{5}{3\left(  2-\lambda\right)
},\sqrt{\frac{2}{3}}\frac{2\lambda+3\lambda^{2}+10\bar{\omega}-16}{3\left(
2-\lambda\right)  }\right), ~~\bar{C}_{2}\left(  0,\frac{2}{3}\sqrt{\frac
{2}{3}}\left(  1+\lambda\right)  \right), 
\]%
\[
\bar{C}_{3}^{\pm}=\left(  \frac{\sqrt{6}\pm\sqrt{3\left(  2-\bar{\omega
}\right)  }}{3\bar{\omega}},\frac{2}{9}\left(  \sqrt{6}\mp2\sqrt{3\left(
2-\bar{\omega}\right)  }\right)  \right), 
\]%
\[
\bar{C}_{4}=\left(  \frac{\sqrt{6}\left(  1+\lambda\right)  }{3\left(
2\bar{\omega}+\lambda-2\right)  },0\right), ~~\bar{C}_{5}^{\pm}=\left(
\frac{\sqrt{6}\pm\sqrt{2\left(  3-2\bar{\omega}\right)  }}{2\bar{\omega}%
},0\right). 
\]
Point $\bar{C}_{2}$ describes the de Sitter solution, $\bar{w}^{\Gamma_{3}%
}\left(  \bar{C}_{2}\right)  =-1$, while the rest of the equilibrium points
describe asymptotic solutions described by an ideal gas with effective
equation of state for the points $\bar{C}_{1}$ and $\bar{C}_{3}^{\pm}$,
describe ideal gas solutions with $\bar{w}^{\Gamma_{3}}\left(  \bar{C}%
_{1}\right)  =\frac{1}{9},$ $\bar{w}^{\Gamma_{3}}\left(  \bar{C}_{3}^{\pm
}\right)  =\frac{1}{9}$,~$\bar{w}^{\Gamma_{3}}\left(  C_{4}\right)
=\frac{10-\lambda-2\lambda^{2}-6\bar{\omega}}{3\left(  2\bar{\omega}%
+\lambda-2\right)  }$ and ~$\bar{w}^{\Gamma_{3}}\left(  C_{5}^{\pm}\right)
=1$. Hence, only $C_{4}$ can describe an accelerated universe when $\left\{
\lambda\leq-2,\bar{\omega}<\frac{4-\lambda^{2}}{2}\right\}  $, $\left\{
-2<\lambda\leq-1,\bar{\omega}<0,\bar{\omega}<\frac{4-\lambda^{2}}{2}\right\}
$, $\left\{  -1<\lambda\leq1,\bar{\omega}<0,\bar{\omega}<\frac{2-\lambda}%
{2}\right\}  $, $\left\{  1<\lambda\leq2,\bar{\omega}<0,0<\bar{\omega}%
<\frac{2-\lambda}{2},\frac{\left(  4-\lambda^{2}\right)  }{2}<\bar{\omega
}<\frac{3}{2}\right\}  $ and $\left\{  \lambda>2,\bar{\omega}<\frac
{4-\lambda^{2}}{2},\frac{2-\lambda}{2}<\bar{\omega}<0,0<\omega<\frac{3}%
{2}\right\}  $.

\begin{figure}[ptbh]
\centering\includegraphics[width=1\textwidth]{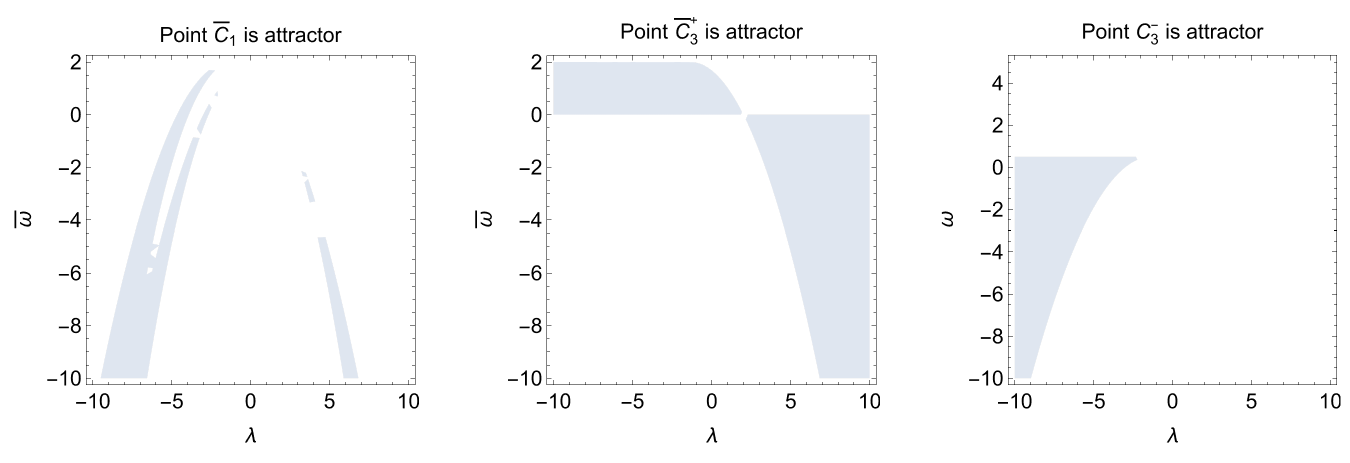}\caption{Region plot in
the space of the free parameters $\left\{  \lambda,\bar{\omega}\right\}  $
where the equilibrium points $\bar{C}_{1}$ and~$\bar{C}_{3}^{\pm}$ are
attractors.}%
\label{ff4}%
\end{figure}

Furthermore, we find that the de Sitter solution described by point $\bar
{C}_{2}$ is always an attractor, while points $\bar{C}_{4}$ is an attractor
when $\left\{  -1<\lambda\leq2,\bar{\omega}<\frac{2-\lambda}{\lambda}%
,2\lambda+3\lambda^{2}+10\bar{\omega}-16>0\right\}  $ or $\left\{  \lambda
\leq-1,\lambda>2,2\lambda+3\lambda^{2}+10\bar{\omega}-16<0,\bar{\omega}%
>\frac{2-\lambda}{\lambda}\right\}  \,$. For the rest of the points, the
regions in the space of the free parameters where the points are attractors
are presented in Fig. \ref{ff4}.

For the study at the infinity, we employ the Poincar\'{e} variables%
\[
\bar{x}=\frac{\bar{X}}{\sqrt{1-\bar{X}^{2}-\bar{\Xi}^{2}}},~\xi=\frac{\bar
{\Xi}}{\sqrt{1-\bar{X}^{2}-\bar{\Xi}^{2}}},~\]
and the new time variable
\[~d\bar{T}=\sqrt{1-\bar{X}^{2}%
-\bar{\Xi}^{2}}d\bar{\tau}.\]
Thus, the dynamical system reads%
\[
\frac{d\bar{X}}{d\bar{T}}=\bar{g}_{1}\left(  \bar{X},\bar{\Xi}\right)
,~~\frac{d\bar{\Xi}}{dT}=\bar{g}_{2}\left(  \bar{X},\bar{\Xi}\right). 
\]

The stationary points at the infinity regime are
\[
\bar{C}_{1\pm}^{\inf}=\left(  0,\pm1\right), ~~\bar{C}_{4\pm}^{\inf}=\left(
\pm1,0\right). 
\]
and%
\begin{align*}
\bar{C}_{2\pm}^{\inf}  &  =\left(  \pm\sqrt{\frac{9\left(  \lambda-1\right)
^{2}}{9\left(  \lambda-1\right)  ^{2}+4\left(  1+\lambda\right)  ^{2}%
\bar{\omega}}},\sqrt{\frac{4\left(  1+\lambda\right)  \bar{\omega}}{9\left(
\lambda-1\right)  ^{2}+4\left(  1+\lambda\right)  ^{2}\bar{\omega}}}\right)
,~~\\
\bar{C}_{3\pm}^{\inf}  &  =\left(  \pm\sqrt{\frac{9\left(  \lambda-1\right)
^{2}}{9\left(  \lambda-1\right)  ^{2}+4\left(  1+\lambda\right)  ^{2}%
\bar{\omega}}},-\sqrt{\frac{4\left(  1+\lambda\right)  \bar{\omega}}{9\left(
\lambda-1\right)  ^{2}+4\left(  1+\lambda\right)  ^{2}\bar{\omega}}}\right). 
\end{align*}

\begin{figure}[ptbh]
\centering\includegraphics[width=1\textwidth]{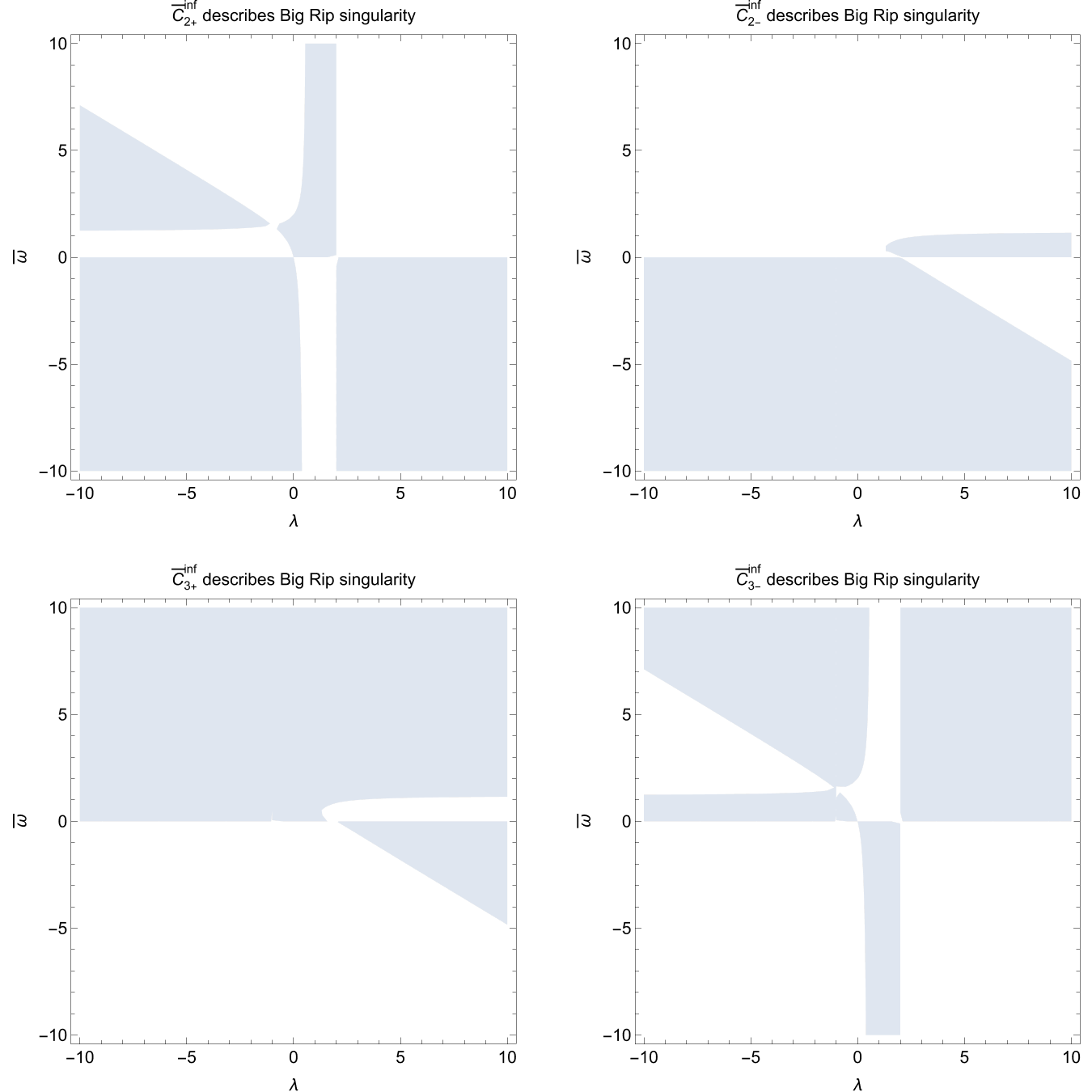}\caption{Region plot in
the space of the free parameters $\left\{  \lambda,\bar{\omega}\right\}  $
where the equilibrium points $\bar{C}_{2\pm}^{\inf}$ and~$\bar{C}_{3\pm}%
^{\inf}$ describe Big Rip singularities.}%
\label{ff5}%
\end{figure}

Points $\bar{C}_{1\pm}^{\inf}$ describe de Sitter solutions, while the new
points $\bar{C}_{4\pm}^{\inf}$ correspond to Big Rip singularities when
$\frac{\left(  \bar{\omega}\left(  \lambda-2+2\bar{\omega}\right)  \right)
}{3-2\bar{\omega}}<0.$ Moreover, points $\bar{C}_{2\pm}^{\inf}$ and $\bar
{C}_{3\pm}^{\inf}$ can describe Big Rip singularities for specific values of
the free parameters. In Fig. \ref{ff5} we present the regions where the
asymptotic solutions at these equilibrium points describe Big Rip singularities.

Finally, we find that points $\bar{C}_{1\pm}^{\inf}$,~$\bar{C}_{2\pm}^{\inf}$
and $\bar{C}_{3\pm}^{\inf}$ describe unstable solutions, while points $\bar
{C}_{4\pm}^{\inf}$ are attractors for $\left\{  \bar{\omega}<0,\lambda
<-1\right\}  $, $\left\{  0<\bar{\omega}<\frac{3}{2},\lambda>2\left(
1-\bar{\omega}\right)  \right\}  $ or $\left\{  \bar{\omega}>\frac{3}%
{2},\lambda<2\left(  1-\bar{\omega}\right)  \right\}  $.

The results of this Section are summarized in Table \ref{tabl3}.

We remark that while in the finite regime, there exists a one-to-one connection
between the stationary points for the two frames, at the infinity regime there
exist a new family of solutions, described by the points $\bar{C}_{4\pm}%
^{\inf}$.

\begin{table}[tbp] \centering
\caption{Equilibrium points and physical properties for the field equations of the third connection for the Jordan and Einstein frames}%
\begin{tabular}
[c]{ccccc}\hline\hline
\textbf{Point} & \textbf{Existence} & $\mathbf{w}^{\Gamma_{3}}$ &
\textbf{Acceleration?} & \textbf{Attractor?}\\\hline
&  &  &  & \\
\multicolumn{5}{c}{\textbf{Equilibrium points for Connection }$\Gamma_{3}%
$\textbf{ in the Jordan frame}}\\\hline
$C_{1}$ & $\lambda\neq\frac{1}{3}$ & $\frac{\lambda-9}{3\left(  3\lambda
-1\right)  }~$ & $\frac{1}{3}<\lambda<2$ & Fig. \ref{ff3}\\
$C_{2}$ & Always & $-1$ & Always & Always\\
$C_{3}^{\pm}$ & $\omega\neq0$ ,$~\omega<\frac{1}{2}$ & $\frac{1}{9}%
+\frac{2\left(  1\pm\sqrt{1-2\omega}\right)  }{9\omega}$ & $-\frac{3}%
{2}<\omega<0$ & Fig. \ref{ff3}\\
$C_{4}$ & $\omega\neq0$ & $1-\frac{1-\lambda}{3\omega}$ & Yes & Yes\\
$C_{5}^{\pm}$ & $\omega<0$ & $1\pm\sqrt{\frac{8}{3\left\vert \omega\right\vert
}}$ & Yes for $C_{5}^{-}$ & Yes for $C_{5}^{-}$\\
$C_{1\pm}^{\inf}$ & Always & $-1$ & Always & No\\
$C_{2\pm}^{\inf}$ & Always & Big Rip & Always & No\\
$C_{3\pm}^{\inf}$ & Always & Big Rip & Always & No\\
&  &  &  & \\
\multicolumn{5}{c}{\textbf{Equilibrium points for Connection }$\Gamma_{3}%
$\textbf{ in the Einstein frame}}\\\hline
$\bar{C}_{1}$ & $\lambda\neq2~$ & $\frac{1}{9}$ & No & Fig. \ref{ff4}\\
$\bar{C}_{2}$ & Always & $-1$ & Always & Always\\
$\bar{C}_{3}^{\pm}$ & $\bar{\omega}\neq0$ ,$~\bar{\omega}<2$ & $\frac{1}{9}$ &
No & Fig. \ref{ff4}\\
$\bar{C}_{4}$ & $\bar{\omega}\neq1-\frac{\lambda}{2}$ & $\frac{10-\lambda
-2\lambda^{2}-6\bar{\omega}}{3\left(  2\bar{\omega}+\lambda-2\right)  }$ &
Yes & Yes\\
$\bar{C}_{5}^{\pm}$ & $\bar{\omega}\neq0$ ,$~\bar{\omega}<\frac{3}{2}$ & $1$ &
No & No\\
$\bar{C}_{1\pm}^{\inf}$ & Always & $-1$ & Yes & No\\
$\bar{C}_{2\pm}^{\inf}$ & $\lambda\neq1$ and$~\bar{\omega}\neq0$ & $\pm
w_{0}\left(  \lambda,\bar{\omega}\right)  \infty$ & Fig. \ref{ff5} & No\\
$\bar{C}_{3\pm}^{\inf}$ & $\lambda\neq1~$and $\bar{\omega}\neq0$ & $\pm
w_{1}\left(  \lambda,\bar{\omega}\right)  \infty$ & Fig. \ref{ff5} & No\\
$\bar{C}_{4\pm}^{\inf}$ & Always & $sign\left(  \frac{\left(  \bar{\omega
}\left(  \lambda-2+2\bar{\omega}\right)  \right)  }{3-2\bar{\omega}}\right)
\infty$ & $\frac{\left(  \bar{\omega}\left(  \lambda-2+2\bar{\omega}\right)
\right)  }{3-2\bar{\omega}}<0$ & Yes\\\hline\hline
\end{tabular}
\label{tabl3}%
\end{table}%

\section{Conclusions}

\label{sec8}

In this study, we investigate the effects of conformal transformations on the
physical properties of solution trajectories in a scalar-nonmetricity
cosmology. Specifically, within the framework of nonmetricity gravity, we
consider a scalar field nonminimally coupled to the Lagrangian of STGR. Our
focus is on the asymptotic dynamics of the field equations, particularly in
the scenario of an isotropic and homogeneous universe described by a spatially
flat FLRW line element.

In General Relativity the Ricci scalar is associated with the
Levi-Civita connection for the metric tensor, while in STGR, the nonmetricity scalar
depends on a connection that is not uniquely defined. We imposed conditions on
the connection to fulfil the symmetries of the background spacetime, be
symmetric and flat, and align with the cosmological principle for a
cosmological fluid. This leads to three families of connections, each
associated with a distinct nonmetricity scalar differing by a boundary term.
Although these connections yield the same limit of field equations in STGR,
the presence of a nonminimally coupled scalar field introduces new
geometrodynamical degrees of freedom related to the boundary term.
Consequently, in scalar-nonmetricity theory, the field equations exhibited
dependence on the choice of connection.

For each of the three cosmological models defined by the different
connections, we analyzed the phase space by identifying equilibrium points and
studying their stability properties. Each equilibrium point corresponds to an
asymptotic solution for cosmological evolution, allowing us to construct the
cosmological history and establish constraints on the free parameters of the
theory. Additionally, we applied the same analysis to the field equations of
three conformal equivalent theories defined in the Einstein frame.

Comparing the physical properties of solutions at equilibrium points for the
three sets of symmetries, we conclude that, regardless of the connection,
there exists a one-to-one relation between equilibrium points in the Jordan
and Einstein frames. Interestingly, the de Sitter universe and solutions
describing Big Rip singularities remained invariant under conformal
transformations. This behavior in nonmetricity gravity contrasts with that in
scalar-curvature theory \cite{c0}, where singular solutions in one frame can
be related to nonsingular solutions in a conformal equivalent theory and vice versa.

The debate over which frame is the \textquotedblleft
physical\textquotedblright\ one was ongoing, see the discussions
\cite{c1,c2,c3,c4,c5,c6,c7,c8}, our study suggests that there are no
significant differences in the cosmological evolution within the context of
nonmetricity gravity. In future research, we plan to extend this investigation
to the case of compact objects.

\begin{acknowledgments}
KD was funded by the National Research Foundation of South Africa, Grant
number 131604. AG was financially supported by FONDECYT 1200293. GL was funded
by Vicerrectoria de Investigacion y Desarrollo Tecnologico (VRIDT) at
Universidad Catolica del Norte through Resoluci\'{o}n VRIDT No. 026/2023 and
Resoluci\'{o}n VRIDT No. 027/2023 and the support of Nucleo de Investigacion
Geometria Diferencial y Aplicaciones, Resoluci\'{o}n VRIDT No. 096/2022. He
also acknowledges the financial support of Proyecto de Investigacion Fondecyt
Regular 2023, Resoluci\'{o}n VRIDT No. 076/2023. AP thanks the support of
VRIDT through Resoluci\'{o}n VRIDT No. 096/2022, Resoluci\'{o}n VRIDT No. 098/2022.
\end{acknowledgments}

\end{document}